\documentclass[aps,prd,reprint,groupedaddress,
 amsmath,amssymb]{revtex4-1}

\usepackage{graphicx}
\usepackage{subfig}
\usepackage[justification=justified]{caption}

\DeclareMathOperator{\sinc}{sinc}

\begin{document}


\title{Gravitational-wave parameter estimation with gaps in LISA: a Bayesian data augmentation method}


\author{Quentin Baghi$^1$}
\email[]{quentin.s.baghi@nasa.gov}
\author{James Ira Thorpe$^1$}
\author{Jacob Slutsky$^1$}
\author{John Baker$^1$}
\author{Tito Dal Canton$^1$}
\author{Natalia Korsakova$^2$}
\author{Nikos Karnesis$^3$}

\affiliation{$^1$NASA Goddard Space Flight Center, 8800 Greenbelt Rd, Maryland, USA}
\affiliation{$^2$Art\'{e}mis UMR 7250, Observatoire de la C\^{o}te d'Azur, Boulevard de l'Observatoire, CS 34229-F, 06304 NICE Cedex 4}
\affiliation{$^3$Laboratoire Astroparticules et Cosmologie, Universit\'{e} Paris Diderot, 10 Rue Alice Domon et L\'{e}onie Duquet, 75013 Paris, France}

%


\date{\today}

\begin{abstract}
By listening to gravity in the low frequency band, between 0.1 mHz and 1 Hz, the future space-based gravitational-wave observatory LISA will be able to detect tens of thousands of astrophysical sources from cosmic dawn to the present. The detection and characterization of all resolvable sources is a challenge in itself, but LISA data analysis will be further complicated by interruptions occurring in the interferometric measurements. These interruptions will be due to various causes occurring at various rates, such as laser frequency switches, high-gain antenna re-pointing, orbit corrections, or even unplanned random events. Extracting long-lasting gravitational-wave signals from gapped data raises problems such as noise leakage and increased computational complexity. We address these issues by using Bayesian data augmentation, a method that reintroduces the missing data as auxiliary variables in the sampling of the posterior distribution of astrophysical parameters. This provides a statistically consistent way to handle gaps while improving the sampling efficiency and mitigating leakage effects. We apply the method to the estimation of galactic binaries parameters with different gap patterns, and we compare the results to the case of complete data.
\end{abstract}


\maketitle

\section{\label{sec:intro}Introduction}

The laser interferometer space antenna (LISA)~\cite{Danzmann2017}, the future space-borne gravitational-wave observatory under development by ESA and NASA, will probe gravitational-wave radiation in the milihertz regime, with a peack sensitivity between 0.1 mHz to 1 Hz. Unlike the ground-based detectors LIGO~\cite{Aasi2015} and Virgo~\cite{Acernese2015}, LISA will be constantly observing a large number of sources, some of them emitting long-lasting signals on periods of months to years. The detection and characterization of all resolvable sources is a challenge for the data analysis~\cite{Babak2010}. In addition, over such time scales, the instrument is very likely to undergo interruptions in the measurement, which will add extra complication in the extraction of gravitational signals.

The LISA observatory is a constellation of three satellites forming a triangle whose side pathlengths are monitored through laser links with a sensitivity level of $\rm pm / \sqrt{Hz}$. Inertial references are provided by free-falling test-masses housed in the satellites, and the gravitational signal is obtained from several interferometric measurements.
In such a complex system, measurement interruptions can be caused by various phenomena. One of them is the re-pointing process of the satellites' high gain antennas, during which the measured data may be saturated or too perturbed to be usable to infer scientific information. Another one is the re-locking of the laser frequencies, which may be needed to maintain heterodyne frequencies in the sensitive bandwidth of the phasemeters. In addition, the measurement is likely to be affected by transient perturbations in the data. Such events have been observed in LIGO-Virgo~\cite{Abbott2017} and in LISA Pathfinder~\cite{Armano2018a} data.
In some cases the safest solution to avoid their impact on the science performance may be to discard the parts of the data where they arise, thereby inducing gaps in the exploitable data streams.

Assessing the impact of data gaps on the observation of gravitational wave sources with LISA is important to quantify the relative impact of different kinds of interruptions, in order to inform the design of the instrument. Furthermore, reducing this impact to minimum is crucial to be able to optimize the scientific return of the mission.

In the problem under study, we define data gaps as the absence of usable data points during certain time spans in time series that are originally evenly sampled. Data gaps can be problematic for two reasons. First, the discrete Fourier transform (DFT) of gapped data is subject to spectral leakage, affecting both the gravitational signal and the stochastic noise. This may lead to increased bias and variance when the computation of the likelihood is done in the Fourier domain. Second, in addition to the leakage effect, another problem is that the diagonal approximation of the covariance matrix in Fourier space is not valid for gapped data. Appropriately weighting the data to account for the noise would require to compute the covariance in the time domain, which is computationally expensive and may not be possible for long data samples. Therefore appropriate analysis methods must be developed.

Few works have addressed the problem of gravitational-wave parameter inference in the presence of data gap. \citet{Pollack2004} studied the recovery of monochromatic signals in the presence of data disturbances and encountered complications for daily gaps. \citet{Carre2010} studied the effect of gaps on the precision of ultra-compact galactic binary (UCB) parameter estimation based on the Fisher information matrix (FIM) approximation. 
Their approach is to apply apodization, i.e. a window in the time domain which has smooth transitions at the gap edges, allowing them to reduce spectral leakage. They provide a first assessment of the impact of hour-long gaps on parameter estimation, showing that errors increase by 2\% to 9\% depending on the parameter. Worst cases are obtained for gap frequencies larger than one per week. 

However, these studies do not entirely address the problem of noise correlations in the presence of gaps. Besides the fact that FIM calculations do not always properly represent correlations between parameters, the diagonal approximation of the noise covariance matrix in the Fourier domain is generally not valid for gapped data. Likewise, treating the remaining data segments as independent measurements may lead to modeling errors. Indeed,  apodization windowing is a sub-optimal weighting of the data, the optimal one being provided by the inverse covariance of the entire vector of observed data.

In the present work we tackle the problem of gaps based on statistical inference, by studying its impact on Bayesian parameter estimation, and by proposing an adapted method that optimally takes into account noise correlations. 

Statistical inference in the presence of missing data is a well covered problem in the statistical literature \cite{Little2002a, Daniels2008, Carpenter2013}. To circumvent the computational issues that can arise when directly computing the likelihood with respect to the observed data, a common trick is to introduce a step where missing data are attributed a statistically consistent value, a process called ``imputation". This allows one to efficiently compute the likelihood from the reconstructed data using standard methods for complete data sets. In the framework of maximum likelihood estimation (MLE), this approach corresponds to the expectation-maximization (EM) algorithm \cite{Dempster1976}. 
In the framework of Bayesian estimation, which is more widely used in gravitational-wave data analysis, the equivalent algorithm is known as data augmentation (DA) \cite{Tanner1987}. In this procedure, the missing data are treated as auxiliary variables of the model, and sampled along with the parameters of interest. The sampling of missing data can be done iteratively, in a block-Gibbs process \cite{Little2002a,P.Murphy1991}. In a way similar to the EM algorithm, the procedure iterates between two steps: an imputation (I) step, where the missing data are drawn from their conditional distribution;  and a posterior (P) step, where the parameters of interest are drawn from their distribution given the current value of the missing data. The algorithm is shown to converge towards the joint posterior distribution of the model parameters and the missing data, given the observed data. 
In this work we implement a data augmentation method that we apply to simulated LISA observations, in order to demonstrate the performance of such an approach.

To analyse measurements of gravitational-wave detectors, we usually model the data by a multivariate Gaussian and stationary distribution \cite{Jaranowski2012}. In this case the conditional distribution of missing data given the observed data is also Gaussian and can be written explicitly. However, it involves the computation of the product of the inverse covariance matrix of observed data with the vector of model residuals. As Fourier diagonalization is not possible, this inversion would require $O(N_o^{3})$ operations, where $N_o$ is the number of observed data points. In the case of stationary noise, more efficient ways to perform this computation are possible, by taking advantage of iterative inversion methods and efficient matrix-to-vector computations using the fast Fourier transform (FFT) \cite{Luo2013,Stroud2014, Baghi2016}. However, we found that this was a too heavy bottleneck for the large number of iterations involved in Markov-chain Monte-Carlo (MCMC) algorithms used to sample the posterior distribution of gravitational-wave parameters. Instead, we adopt an approximation to perform the imputation step, which is done conditionally on the nearest observations around gaps \cite{Datta2016}. 

In this work we develop a blocked-Gibbs data augmentation algorithm which estimates the posterior distribution of signal and noise parameters. In order to demonstrate the performance of this method on a simple example, we apply it to the characterization of UCBs in simulated LISA data. In Sec.~\ref{sec:modeling} we present the general Gaussian stationary model that we adopt to describe gravitational-wave measurements, both in the complete and gapped data cases. Then in Sec.~\ref{sec:windowing} we describe the standard method of time-domain windowing that can be used to handle data gaps. We show how to optimize it before highlighting its drawbacks. This leads us to introduce the data augmentation method as an alternative approach in Sec.~\ref{sec:da}, where we describe its theoretical basis. In Sec.~\ref{sec:case_study} we present an application to the case of UCB parameter estimation, where we describe the time-domain model used to simulate LISA data and the frequency-domain model used for the data analysis. In Sec.~\ref{sec:simulations} we detail the simulations used in this study, and in Sec.~\ref{sec:estimation_results} we present the results of the gravitational-wave parameter estimation. We finally draw conclusions in Sec.~\ref{sec:conclusions}.

\section{\label{sec:modeling}General statistical model}

In this section we introduce the model used to describe the data, both in the case of complete and gapped data series.

\subsection{Complete data case}

In gravitational-wave astronomy, interferometric measurements are time series $\boldsymbol{y}$ sampled at some frequency $f_s$. Temporarily putting aside the physical quantity that they represent, they can generally be modeled as the sum of a gravitational signal and a noise term:
\begin{eqnarray}
\label{eq:model}
\boldsymbol{y} = \boldsymbol{h}\left(\boldsymbol{\theta}_{h}\right) + \boldsymbol{n}\left(\boldsymbol{\theta}_{\mathrm{n}}\right),
\end{eqnarray}
where $\boldsymbol{y}$ is a $N \times 1$ vector containing the measured data points, $\boldsymbol{h}$ is the signal due to the incoming gravitational waves (GW) depending on a vector of parameters $\boldsymbol{\theta}_{h}$. The vector $\boldsymbol{n}$ represents the random measurement noise and is a zero-mean, stationary Gaussian random variable of covariance $\boldsymbol{\Sigma}$ with probability density function
\begin{eqnarray}
\label{eq:noise_pdf}
p\left(\boldsymbol{n}\right) = \frac{1}{ \sqrt{ (2 \pi)^{N} \left| \boldsymbol{\Sigma} \right|} } \exp\left\{ -\frac{1}{2} \boldsymbol{n}^{T} \boldsymbol{\Sigma}^{-1}\boldsymbol{n}\right\}.
\end{eqnarray}
For a complete, evenly sampled time series, the covariance of the noise has a Toeplitz structure, which means that its elements are constant along diagonals. They are directly given by the autocovariance function $R(t)$, which is related to the power spectral density (PSD) function $S_n(f)$ such that $\forall (p,q) \in \left[0, \, N-1\right]^{2}$,
\begin{eqnarray}
\label{eq:cov_from_psd}
\Sigma(p,q) = R\left( \frac{p - q}{f_s} \right) =\int_{-\frac{f_s}{2}}^{+\frac{f_s}{2}} S_{n}(f) e^{2j\pi f \frac{p-q}{f_s} } df.
\end{eqnarray}
The PSD can be parametrized by some parameter vector $\boldsymbol{\theta}_{\mathrm{n}}$. 
For a sufficiently large number of points $N$, a Toeplitz matrix can be approximated by a circulant matrix, which is diagonalizable in the Fourier basis:
\begin{eqnarray}
\label{eq:diagonal_fourier}
\boldsymbol{\Sigma} \approx \boldsymbol{F}_{N}^{*} \boldsymbol{\Lambda} \boldsymbol{F}_{N},
\end{eqnarray}
where $\boldsymbol{F}_{N}$ is the discrete Fourier transform matrix defined as $F_{N}(l,m) = N^{-1/2} \exp{\left( - \frac{2 \pi j l m}{N} \right)}$, where we $j = \sqrt{-1}$ denotes the complex number. $\boldsymbol{\Lambda}$ is a diagonal matrix whose elements are directly related to the PSD as $\Lambda_{kk} = f_s S\left(f_k\right)$, where $f_k$ are the frequency elements of the Fourier grid $f_k = f_s\frac{k}{N}  \text{ if } 0 \leq k \leq \lfloor \frac{N-1}{2} \rfloor$, and $f_k = -f_s\frac{N-k}{N}$ otherwise. 


Up to a constant, the log-likelihood for model (\ref{eq:model}) with a noise distribution given by Eq.~(\ref{eq:noise_pdf}) can be written as
\begin{eqnarray}
\label{eq:gaussian_like}
\log p\left( \boldsymbol{y} | \boldsymbol{\theta} \right) = - \frac{1}{2} \left[ \log \left| \boldsymbol{\Sigma} \right| +  \left(\boldsymbol{y} - \boldsymbol{h} \right)^T \boldsymbol{\Sigma}^{-1} \left(\boldsymbol{y} - \boldsymbol{h} \right)  \right],
\end{eqnarray}
where $\boldsymbol{\theta}  \equiv \begin{pmatrix}  \boldsymbol{\theta}_{h}^{T} & \boldsymbol{\theta}_{\mathrm{n}}^{T} \end{pmatrix}^{T}$ gathers all the parameters describing the model, i.e. both signal and noise.

Eq.~(\ref{eq:diagonal_fourier}) allows us to write the model log-likelihood using the Whittle's approximation \cite{Whittle1954}:
\begin{eqnarray}
\label{eq:whittle_like}
\log p\left( \boldsymbol{y} | \boldsymbol{\theta} \right) \approx - \frac{1}{2} \sum_{k=0}^{N-1} \left[ \log \Lambda_{kk}  + \frac{\left| \tilde{y}_k - \tilde{h}_k \right|^{2} }{\Lambda_{kk} }  \right],
\end{eqnarray}
where for any vector $\boldsymbol{x}$, the notation $\boldsymbol{\tilde{x}}$ designates its discrete Fourier transform:
\begin{eqnarray}
\boldsymbol{\tilde{x}}  \equiv  \boldsymbol{F}_{N} \boldsymbol{x}. 
\end{eqnarray}
The likelihood in Eq.~(\ref{eq:whittle_like}) can be efficiently computed since it only involves element-wise operations on vectors. In addition, if the signal is narrow-banded it can be restricted to a short frequency interval, so that the sum involves a small number of elements.

\subsection{\label{seq:modeling_miss}Modeling missing data}
Let us now introduce the possibility to have some gaps in the time series. Gaps are identified by a mask $\boldsymbol{w}$ such that $w(n)=0$ if data $n$ is unavailable, and $w(n) \in  \left] 0,\, 1\right]$ if data $n$ is observed (values may be lower than 1 for smoothing). In the following, subscripts $o$ and $m$ respectively mean ``observed" and ``missing". Let $N_o$ be the number of observed data points, and $N_m = N - N_o$ the number of missing data points. If we label $i_o(q) \; \forall q \in \left[ 0 \, , \, N_o - 1 \right] $ the indices of observed data, we can form the observed data vector $\boldsymbol{y}_o$ such that $y_{o}(q) = y_{i_o(q)}$. We can likewise define the missing data vector $\boldsymbol{y}_{m}$ of size $N_m$. 
It is also formally useful to define the matrix operator $\boldsymbol{W}_{o}$ (respectively $\boldsymbol{W}_{m}$) that maps the complete data vector to the observed data vector (respectively the missing data vector):
\begin{eqnarray}
\boldsymbol{y}_{o} &=& \boldsymbol{W}_{o}  \boldsymbol{y} ; \nonumber \\
\boldsymbol{y}_{m} &=& \boldsymbol{W}_{m}  \boldsymbol{y}.
\end{eqnarray}
Using this notation, the covariance of $\boldsymbol{y}_o$ with itself is given by $\boldsymbol{\Sigma}_{oo} = \boldsymbol{W}_{o} \boldsymbol{\Sigma} \boldsymbol{W}_{o}^{T}$. Unlike $\boldsymbol{\Sigma}$, the matrix $\boldsymbol{\Sigma}_{oo}$ is not Toeplitz, and approximations (\ref{eq:diagonal_fourier}) and (\ref{eq:whittle_like}) do not hold anymore. As a result, in principle the likelihood should be computed using Eq.~(\ref{eq:gaussian_like}),  replacing $\boldsymbol{y}$  by $\boldsymbol{y}_o$ and $\boldsymbol{\Sigma}$ by $\boldsymbol{\Sigma}_{oo}$. However, the computational cost of $\boldsymbol{\Sigma}_{oo}^{-1} \boldsymbol{z}$ for any vector $\boldsymbol{z}$ can be prohibitive.
To avoid this, we ideally want to come back to a situation similar to the complete data case, and use the convenient formulation of Whittle's likelihood in Eq.~(\ref{eq:whittle_like}). Approaches following this idea are described below.

\section{\label{sec:windowing}The windowing approximation method}

One rather straightforward way to deal with data gaps is to make the assumption that the Fourier transform of the masked data is approximately equal to the Fourier transform of the complete data, which is acceptable if the signal is stationary or has a short frequency bandwidth. Formally, if we define the masked data vector $\boldsymbol{y}_w$ as $y_{w}(i) = w(i) y(i)$, then we can assume that the DFTs of $\boldsymbol{y}_w$ and $\boldsymbol{y}$ are equal up to a normalizing constant:
\begin{eqnarray}
\label{eq:windowing_approx}
\sqrt{r(w)} \boldsymbol{\tilde{y}}_w & \approx &  \boldsymbol{\tilde{y}},
\end{eqnarray}
where we have applied an extra factor $\sqrt{r(w)}$ to the DFT, where $r(w) = N / N_{w} $ and $N_w \equiv \sum_{n=0}^{N-1} w_n$ in order to take into account the loss of power due to masking.
Note that for this approximation to be valid, the mask vector $\boldsymbol{w}$ must be smooth enough, gradually going to zero at the gap edges, in order to limit frequency leakage as much as possible. This is the approach adopted in Ref.~\cite{Carre2010}, and is referred to as the ``windowing method".

One could argue that we could also treat gapped data as a sequence of independent segments. We could then write a likelihood (\ref{eq:whittle_like}) for each of them, and approximate the full likelihood by the sum of the individual segment likelihoods. While this approach could be considered in the case of a few large gaps, it rises two problems when it comes to frequent and relatively short gaps as we assume here. First, short and frequent segments may happen within the noise autocorrelation time and thus cannot be treated as long and widely separated segments, i.e. independent experiments. Second, this would result in a loss of frequency resolution that can be problematic when estimating low-frequency signals with short bandwidths. Therefore we restrict our baseline approach to the windowing method.

\subsection{\label{sec:leakage}Quantification of leakage}

In spite of smooth tapering, using the windowing approximation induces a residual leakage which affects both signal and noise. We quantitatively study this leakage effect in this section.

Let $\tilde{\boldsymbol{h}}(f)$ be the discrete Fourier transform of the GW signal $\boldsymbol{h}$ in Eq.~(\ref{eq:model}). 
For simplicity, here we approximate the DFT of vectors $\boldsymbol{h}$ and $\boldsymbol{w}$ by the continuous Fourier transforms of the GW response function $h(t)$ and the mask function $w(t)$, that we respectively label $H(f)$ and $W(f)$. 
Then the DFT of the masked GW signal can be approximated by the convolution of the original signal Fourier transform with the mask Fourier transform:
\begin{eqnarray}
\label{eq:signal_leakage}
H_{w}(f_k) = \int_{-f_s/2}^{+f_s/2} H(f') W(f_k - f') df',
\end{eqnarray}
where $f_k$ is the $\rm k^{th}$ element of the Fourier grid.
This expression illustrates the intuitive fact that the broader the Fourier transform of the window, the larger the error from masking. In addition, the shorter the signal bandwidth, the smaller the error since $H(f)$ will quickly drop to zero far from its central frequency $f_0$.


The noise is also affected by windowing in time domain. One can show~\cite{priestley1982spectral,Baghi2015} that the diagonal elements of the covariance of the windowed noise DFT $\boldsymbol{\tilde{n}}_w$ are given by the convolution of the true spectral density with the mask periodogram:
\begin{eqnarray}
\label{eq:noise_leakage}
S_{wn}(f) \approx \int_{-\frac{f_s}{2}}^{+\frac{f_s}{2}} S_n(f - f') \left| \tilde{w}(f') \right|^2 df',
\end{eqnarray}
where $\tilde{w}(f) = N^{-1/2} \sum_{n=0}^{N-1} w(n) e^{-2j\pi n f/f_s}$ is the DFT of the mask calculated at frequency $f$. 
For what follows, it is useful to determine a figure of merit quantifying the amount of noise power leakage that affects the estimation. 
To this end, we use the expression of the signal-to-noise ratio (SNR) that we adapt to take into account the leakage. 
In the absence of gaps, the general, continuous approximation formula adopted in GW analysis is \cite{Jaranowski2012}:
\begin{eqnarray}
\rho \equiv 4 \int_{0}^{\infty} \frac{\left| H(f)\right|^2}{S_{1n}(f)}df,
\end{eqnarray}
where $S_{1n}(f) = 2 S_{n}(f)$ is the one-sided noise PSD. 
It is then sensible to define an effective SNR in the presence of masking:
\begin{eqnarray}
\label{eq:SNR_eff}
\rho_{w} \equiv 4 \int_{0}^{\infty} \frac{\left| H_{w}(f)\right|^2}{S_{1wn}(f)}df,
\end{eqnarray}
where $H_{w}(f)$ is given by Eq.~(\ref{eq:signal_leakage}) and $S_{1wn}(f) = 2S_{wn}(f)$ with $S_{wn}(f)$ given by Eq.~(\ref{eq:noise_leakage}). Note that the effective SNR depends on the source, the noise PSD, the integration time and the mask pattern. We will see that the convolution effects involved can degrade the SNR with respect to the complete data case. That is why we propose an alternate approach to the windowing method, that allows us to avoid any leakage effect.

\section{\label{sec:da}A data augmentation method}

The alternative presented in this study is to treat missing data as auxiliary variables to be estimated as part of the parameter estimation scheme. We detail such a strategy in this section.

\subsection{Iterative blocked-Gibbs sampler}

The purpose of any Bayesian estimation method is to estimate the posterior probability density of the parameters of interest, given the data at hand. Formally, this is expressed by Bayes' theorem 
\begin{eqnarray}
\label{eq:bayes}
p\left(\boldsymbol{\theta}|\boldsymbol{y} \right) = \frac{p\left( \boldsymbol{y} | \boldsymbol{\theta}\right) p(\boldsymbol{\theta}) }{\int_{\Theta} p\left( \boldsymbol{y} | \boldsymbol{\theta}\right) p(\boldsymbol{\theta}) d\boldsymbol{\theta} },
\end{eqnarray}
where $p\left(\boldsymbol{\theta}|\boldsymbol{y} \right)$ is the posterior distribution of the parameters, $p\left( \boldsymbol{y}| \boldsymbol{\theta}\right)$ is the likelihood function, $p(\boldsymbol{\theta})$ is the prior distribution and the denominator is the evidence which acts as a normalizing constant.

When there are data gaps, our goal is to compute the posterior distribution given the observed data only. 
This means that we want to sample $p\left(\boldsymbol{\theta}|\boldsymbol{y}_{o} \right)$ using Eq.~(\ref{eq:bayes}). While the complete-data likelihood can be efficiently computed using Eq.~(\ref{eq:whittle_like}), the gapped-data likelihood $p\left( \boldsymbol{y}_o| \boldsymbol{\theta}\right)$ is usually hard to compute, because of the reasons mentioned in Sec.~\ref{seq:modeling_miss}. Thus we need a workaround to use the computational convenience of the complete data-likelihood while properly probing the gapped-data likelihood. This is done by data augmentation (DA) \cite{Tanner1987}, also called Bayesian multiple imputation. As mentioned in Sec.~\ref{sec:intro}, this algorithm iterates between two steps. The first one is the I-step, where the missing data vector is drawn conditionally on the observed data and on the current state of the parameters. The second one is the P-step, where the parameter vector $\boldsymbol{\theta}$ is drawn from the posterior distribution given both the observed and missing data, exactly as in standard Bayesian estimation. 
Thus, in the DA algorithm the update from iteration $i$ to iteration $i+1$ is done as:
\begin{eqnarray}
\label{eq:da_algorithm}
\text{I-step: } && \text{draw } \boldsymbol{y}_{m}^{(i+1)} \sim p\left( \boldsymbol{y}_{m} | \boldsymbol{y}_{o} , \boldsymbol{\theta}^{(i)} \right) \, ;\nonumber \\
\text{P-step: } && \text{draw } \boldsymbol{\theta}^{(i+1)}  \sim p\left( \boldsymbol{\theta} | \boldsymbol{y}_{o} , \boldsymbol{y}_{m}^{(i+1)} \right).
\end{eqnarray}
This scheme corresponds to a blocked Gibbs sampler, because two blocks of parameters ($\boldsymbol{y}_m$ and $\boldsymbol{\theta}$) are drawn sequentially, conditionally to the value of the other parameter at previous iteration. In the following we detail the two steps of the algorithm.

\subsection{The imputation step}

The imputation step draws a value for the missing data vector $\boldsymbol{y}_m$ from its posterior distribution, given the current value of the model parameters and the observed data. If the data is assumed to follow a multivariate Gaussian distribution, then the conditional distribution $p\left( \boldsymbol{y}_m | \boldsymbol{y}_o , \boldsymbol{\theta} \right)$ is also Gaussian, with mean and covariance given by
\begin{eqnarray}
\label{eq:miss_conditional}
\boldsymbol{\mu}_{m|o} & = & \boldsymbol{h}_{m}(\boldsymbol{\theta}) + \boldsymbol{\Sigma}_{mo} \boldsymbol{\Sigma}_{oo}^{-1} \left( \boldsymbol{y}_{o} - \boldsymbol{h}_{o}(\boldsymbol{\theta}) \right) ;\\
\boldsymbol{\Sigma}_{m|o} & = & \boldsymbol{\Sigma}_{mm} - \boldsymbol{\Sigma}_{mo}  \boldsymbol{\Sigma}_{oo}^{-1}  \boldsymbol{\Sigma}_{mo}^{T},
\end{eqnarray}
where $\boldsymbol{\Sigma}_{mm} \equiv \boldsymbol{W}_{m} \boldsymbol{\Sigma} \boldsymbol{W}_{m}^{T}$ is the covariance of the missing data vector, and $\boldsymbol{\Sigma}_{mo} \equiv \boldsymbol{W}_{m} \boldsymbol{\Sigma} \boldsymbol{W}_{o}^{T}$ is the covariance between the missing and the observed data vectors. 
Eq.~(\ref{eq:miss_conditional}) involves the inverse of the covariance matrix $\boldsymbol{\Sigma}_{oo}$. A direct computation of this matrix would be cumbersome for large data sets (i.e. for $N_o > 10^{3}$). Instead, it is possible to iteratively solve the system $\boldsymbol{\Sigma}_{oo} \boldsymbol{z} = \boldsymbol{b}$ for $\boldsymbol{z}$ using the preconjugate gradient algorithm, because the matrix-to-vector product $\boldsymbol{\Sigma}_{oo} \boldsymbol{z}$ can be efficiently computed using fast Fourier transform (FFT) algorithms \cite{Fritz2009,Stroud2014}. 
However, in our application we found that the number of iterations needed to reach a reasonable precision is still restrictive, since a large number of imputation steps must be done to sample the posterior distribution of the parameters.

As an alternative, we make the assumption that the conditional distribution of any missing data $y_m(i)$ mainly depends on the nearest observed points, which is true for fast decaying autocovariance functions. More particularly, let us consider a data gap that we label $j$. We denote $\boldsymbol{y}^{(j)}_{m}$ the vector of data lying inside that gap. Now let $\boldsymbol{y}^{(j)}_{o}$ be a subset of observed data $\boldsymbol{y}_{o}$ which includes the nearest points to gap $j$ (previous and following the gap). Then we can assume that 
\begin{eqnarray}
\label{eq:nn_approx}
p\left( \boldsymbol{y}^{(j)}_{m} | \boldsymbol{y}_{o} \right) \approx p\left( \boldsymbol{y}^{(j)}_{m} | \boldsymbol{y}^{(j)}_{o}\right).
\end{eqnarray} 
In the following, the subset $\boldsymbol{y}^{(j)}_{o}$ is defined as the union of the $N_{j}$ available data points before gap $j$ and the $N_{j}$ data points coming right after. Then drawing a realization of the conditional distribution defined by Eqs.~(\ref{eq:miss_conditional}) has a complexity in $O\left(N_j^2\right)$, hence$N_{j}$ must be kept as small as possible.

\subsection{The posterior step}

In the posterior step we assume that the missing data vector $\boldsymbol{y}_{m}$ is given. Then model parameters $\boldsymbol{\theta}$ can be drawn from the posterior distribution $p\left( \boldsymbol{\theta} | \boldsymbol{y}_{o} , \boldsymbol{y}_{m}\right) $.
To that end, we use a Metropolis-Hastings step: if we denote $\boldsymbol{\theta}^{(i)}$ the value of the parameters at the previous iteration, we use a probability density $q\left( \boldsymbol{x}^{\prime} | \boldsymbol{x} \right) $ to propose a new value $\boldsymbol{\theta}_0$ for the parameters. 
We then accept this proposal with a probability given by the Metropolis ratio
\begin{eqnarray}
A\left(\boldsymbol{x}^{\prime} , \boldsymbol{x} \right) \equiv \mathrm{min}\left\{ 1 \, , \, \frac{ p\left( \boldsymbol{x}^{\prime} | \boldsymbol{y} \right) q\left( \boldsymbol{x} | \boldsymbol{x}^{\prime} \right)  }{ p\left( \boldsymbol{x} | \boldsymbol{y} \right) q\left( \boldsymbol{x}^{\prime} | \boldsymbol{x}  \right) }   \right\} ,
\end{eqnarray}
that we calculate for $\boldsymbol{x}' = \boldsymbol{\theta}_0,\, \boldsymbol{x} = \boldsymbol{\theta}^{(i)}$. 

For our purpose, it will be convenient to separate the update of the model parameters in two Gibbs sub-steps as done by \citet{Edwards}, where we first update the noise parameters $\boldsymbol{\theta}_n$ and then the signal parameters $\boldsymbol{\theta}_h$:
\begin{eqnarray}
\label{eq:posterior_steps}
\text{P1: } \boldsymbol{\theta}_{n}^{(i+1)} & \sim & p\left(  \boldsymbol{\theta}_{n} | \boldsymbol{\theta}_{h}^{(i)} , \boldsymbol{y}_{o} , \boldsymbol{y}_{m}^{(i+1)} \right) \nonumber \\
\text{P2: } \boldsymbol{\theta}_{h}^{(i+1)} & \sim & p\left(  \boldsymbol{\theta}_{h} | \boldsymbol{\theta}_{n}^{(i)} , \boldsymbol{y}_{o} , \boldsymbol{y}_{m}^{(i+1)} \right).
\end{eqnarray}
As the noise PSD does not depend on $\boldsymbol{\theta}_{h}$ and the GW signal does not depend on $\boldsymbol{\theta}_{n}$, their conditional distributions are well separable, making this scheme efficient to perform. 

\section{\label{sec:case_study}Case study: the example of compact galactic binaries}

In this section we describe the model adopted for the gravitational-wave signal $\boldsymbol{h}$, as well as the noise $\boldsymbol{n}$. We restrict our analysis to the case of non-merging, slowly chirping UCBs. Since the aim of this paper is to assess and minimize the impact of data gaps on the LISA science performance, this choice is motivated by the relative simplicity of the signal model. 
In the following, we differentiate between the model that we use to generate the synthetic data set (the ``simulation model"), and the model that we use for the parameter estimation (the ``data analysis model").

\subsection{Simulation model}

\subsubsection{\label{sec:sim_gw_model}Time domain model for gravitational-wave signal}

Let us consider a source located by radius $r$, colatitude $\theta$ and longitude $\phi$ in the solar-system barycentric ecliptic coordinate system. We assume that this source emits a gravitational wave with strain polarizations $h_{+}$ and $h_{\times}$ in the source frame. We consider one of the interferometer arms of the LISA constellation, labeled $i$, whose direction is given by a unit vector $\boldsymbol{n}_i$, pointing towards the receiving spacecraft, and whose length is given by $L_{i}$. Then the incoming wave on the detector will induce a optical phase shift in the laser link, which can be written as \cite{Cutler1998,Cornish2003}
\begin{eqnarray}
\label{eq:phase_shift_arm}
\Delta \Phi^{(i)} (t) = h_{+}(t - d ) F^{(i)}_{+}(t) + h_{\times}(t - d ) F^{(i)}_{\times}(t),
\end{eqnarray}
where the functions $F_{+}$, $F_{\times}$ account for the time-varying projection of the metric components onto LISA's interferometer arms. They depend on sky location $(\theta,\phi)$ and polarization $\psi$ angles of the source, and are better detailed in Appendix~\ref{sec:modulation_functions}. 
The time delay $d$ corresponds to the time-dependent projection of the wave vector onto the vector $\boldsymbol{r}_0$ of modulus $R$ localizing the barycenter of LISA's constellation from the solar-system barycenter (SSB) and reads:
\begin{equation}
d(t) \equiv -\frac{R \sin \theta}{c} \cos\left( \phi - \Phi_T(t) \right).
\end{equation}
This delay depends on the orbital angular position located from the SSB, that we approximate by $\Phi_T(t) \approx 2 \pi t /T $, where $T$ is LISA's orbital period (1 year). 
In the Newtonian limit the gravitational wave polarizations can be written as
\begin{eqnarray}
\label{eq:polarizations}
h_{+}(t) & = &  h_{0+}(t) \cos\left(\Phi_s(t)  + \phi_0 \right); \\ \nonumber 
h_{\times}(t) & = & - h_{0\times}(t) \sin\left(\Phi_s(t) + \phi_0 \right),
\end{eqnarray}
where $h_{0+}(t)$ is generally a time-varying amplitude and $\Phi_s(t)$ is the phase of the gravitational wave.

For sufficiently small binary masses and frequencies, the amplitudes and phases of the above model can be approximated by:
\begin{eqnarray}
 h_{0+} & = & \frac{h_0}{2} \left( 1 + \cos^2 i \right); \nonumber \\
 h_{0\times} & = &  h_0 \cos i ,
\end{eqnarray}
where $i$ is the inclination of the source orbital plane with respect to the direction of propagation of the incoming wave. 
The phase is modeled to the first post-Newtonian order \cite{Blanchet2002}, which we approximate up to second order in time:
\begin{eqnarray}
\Phi_s(t) = 2 \pi f_0 t + \pi \dot{f}_0 t^{2},
\end{eqnarray}
where $\dot{f}_0$ is the time derivative of the frequency, assumed constant. 

In the case of LISA, the observables of interest are given by time delay interferometry (TDI) rather than the phasemeter measurement themselves, in order to deal with the fact that we have unequal and time-varying armlengths. They are constructed from a delayed linear combination of phasemeter measurements tailored to cancel the otherwise overwhelming laser frequency noise~\cite{Tinto2014}. Assuming equivalence of clockwise and counterclockwise light propagation direction, the first generation TDI Michelson observable $X$ is given by \cite{Armstrong1999,Otto2015}:
\begin{eqnarray}
\label{eq:TDI1}
X_{1} & = &  \left( \Delta \Phi_{2':\boldsymbol{322'}}  + \Delta \Phi_{1:\boldsymbol{22'}} + \Delta \Phi_{3:\boldsymbol{2'}} + \Delta \Phi_{1'} \right)  \nonumber \\
&& - \left(  \Delta \Phi_{3:\boldsymbol{2'3'3}} - \Delta \Phi_{1':\boldsymbol{3'3}} + \Delta \Phi_{2':\boldsymbol{3}} + \Delta \Phi_{1} \right) ,
\end{eqnarray}
where the colon indicate the application of a delay operator 
\begin{eqnarray}
f(t)_{: \boldsymbol{k}} = f\left( t - \frac{L_k}{c}\right).
\end{eqnarray}
The primes in Eq.~(\ref{eq:TDI1}) indicate that the delay is taken in the counterclockwise direction. In this study we assume that all arms have equal and constant lengths, but we use Eq.~(\ref{eq:TDI1}) to account for the effect of TDI on noise correlations and signal response. Note that other TDI variables can be obtained from similar combinations.

\subsubsection{\label{sec:sim_noise_model}Noise model}

The noise model used to generate the data in this study is specified in Ref.~\cite{Stebbins}, and can be written in relative frequency units as:
\begin{eqnarray}
\label{eq:sim_psd_model}
S_{X}(f) & = & 16  \sin^2\left(\frac{f}{f_{*}}\right) \Bigg[2  \left(1 + \cos^2\left(\frac{f}{f_{*}}\right) \right)  S_{a\nu}(f) \nonumber \\
&&   + S_{o\nu}(f) \Bigg],
\end{eqnarray}
where $f_{*} \equiv \frac{c}{2 \pi L} \approx 19$ mHz is the LISA response transfer frequency and $S_{a\nu}(f)$ and $S_{o\nu}(f)$ are respectively the test-mass acceleration and the optical metrology system noise PSDs:
\begin{eqnarray}
S_{a\nu} &=& 10^{-30} \left[1 + \left( \frac{f_{a1}}{f} \right)^{2} \right]\left[1+\left(\frac{f}{f_{a2}}\right)^4\right] \left(2 \pi f c \right)^{-2} ; \nonumber \\
S_{o\nu} &=& 2.25 \cdot 10^{-22}   \left[ 1 + \left( \frac{f_{o}}{f} \right)^{4} \right] \left( \frac{2 \pi f}{c} \right)^2,
\end{eqnarray}
where the inflection frequencies are $f_{a1} = 4 \cdot 10^{-4}$ Hz, $f_{a2} = 8\cdot 10^{-3}$ Hz and $f_{o} = 2 \cdot 10^{-3}$ Hz. The resulting PSD shape is a convex function of frequency reaching its minimum at about 2 mHz, with a low-frequency slope of $f^{-4}$ and a high-frequency slope of $f^{2}$, modulated by the LISA response and the effect of TDI delays (see Fig.~\ref{fig:psd_estimation}).


\subsection{Data analysis model}

\subsubsection{Frequency domain model for gravitational-wave signal}

In order to efficiently compute the likelihood in Eq.~(\ref{eq:whittle_like}), it is preferable to have an analytic model of the response in the frequency domain. This model can be approximately derived in two steps. 

First, in the case of model (\ref{eq:polarizations}) one can show that Eq.~(\ref{eq:phase_shift_arm}) can be re-written in the form of a linear combination of oscillating functions \cite{Krolak2004,Cornish2005,Baut2009} as 
\begin{eqnarray}
\label{eq:phase_linear_time_model}
\Delta \Phi^{(i)} (t) = \sum_{j=1}^{4} a^{(i)}_{j} g_{j}(t).
\end{eqnarray}
The amplitudes $a_{j}^{(i)}$ only depend on the source's sky localization and orientation angles, as well as its amplitude. The  functions $g_{j}(t)$ (whose expressions are given in Appendix~\ref{sec:fourier_series}) depends on the source's frequency and frequency derivative, and has a delay term depending on its sky location. The parameters characterizing $g_{j}$ are usually called intrinsic parameters. 

If the frequency of the incoming wave is small with respect to the interspacecraft travel time $L/c$ (low frequency approximation), the TDI combination written in Eq.~(\ref{eq:TDI1}) acts like a differential operator and the TDI response can similarly be written as a linear combination:
\begin{eqnarray}
\label{eq:tdi_linear_time_model}
X_{1} (t) = \frac{4 L }{c}\sum_{k=1}^{4} a_{Xk} \dot{g}_{k}(t) ,
\end{eqnarray}
where we set $a_{Xj} = a_{2j} - a_{3j}$. 
In the following, we express TDI in relative frequency shift $\delta \nu / \nu_0$, which is obtained from the phase shift by applying a time derivation and dividing by the laser frequency. Finally, in the frequency domain the waveform can be written as \cite{Cornish,Bouffanais2016}
\begin{eqnarray}
\label{eq:TDI_frequency}
\tilde{X}_{\nu} (f) = \sum_{k=1}^{4} a_{Xk} \tilde{g}_{\nu k}(f), 
\end{eqnarray}
where we defined the function $\tilde{g}_{\nu}(f) \equiv -\frac{8 \pi f^2 L }{\nu_0 c} \tilde{g}(f)$. 
The formulation in Eq.~(\ref{eq:TDI_frequency}) is useful for data analysis, because it can be converted into matrix notation as
\begin{eqnarray}
\boldsymbol{\tilde{h}} = \boldsymbol{\tilde{M}} \boldsymbol{a},
\end{eqnarray}
where $\boldsymbol{\tilde{h}}$ is the signal DFT vector with elements $\tilde{h}(p) = \tilde{X}_{\nu} (f_p) $, $\boldsymbol{a}$ is the amplitude vector with elements $a(k) = a_{Xk}$ and $\boldsymbol{\tilde{M}}$ is a design matrix with elements $\tilde{M}(p,k) = \tilde{g}_{\nu k}(f_p)$. As a result, sampling the GW parameters (i.e. performing step P1 of the DA algorithm in Eq.~(\ref{eq:posterior_steps})) can be done in two Gibbs steps: (i) sample for the intrinsic parameters using a Metropolis-Hastings step, and (ii) sample for $\boldsymbol{a}$ using its conditional distribution, which is a Gaussian distribution whose mean and variance can be written explicitly:
\begin{eqnarray}
\label{eq:extrinsic_distrib}
\boldsymbol{a} & \sim & \mathcal{N}\left( \boldsymbol{\mu}_a , \, \boldsymbol{C}_a \right) ; \nonumber \\
\boldsymbol{\mu}_a & = & \left( \boldsymbol{\tilde{M}}^{*} \boldsymbol{\Lambda}^{-1} \boldsymbol{\tilde{M}} \right)^{-1}  \boldsymbol{\tilde{M}}^{*} \boldsymbol{\Lambda}^{-1}   \boldsymbol{\tilde{y}} ; \nonumber \\
\boldsymbol{C}_a & = &  \left( \boldsymbol{\tilde{M}}^{*} \boldsymbol{\Lambda}^{-1} \boldsymbol{\tilde{M}} \right)^{-1}.
\end{eqnarray}
This scheme allows us to set up a MCMC algorithm which reduces to~4 parameters ($\theta,\, \phi, \, f_0, \, \dot{f}_0$) instead of~8  ($h_0, \, \phi_0, \, i, \psi, \, \theta,\, \phi, \, f_0, \, \dot{f}_0$). This is similar to using the $\mathcal{F}$-statistics in the search phase, where one marginalizes over parameters $\boldsymbol{a}$~\cite{Jaranowski2012}. Once the GW parameters are updated, we form the model residuals $\boldsymbol{\tilde{y}} - \boldsymbol{\tilde{M}} \boldsymbol{a}$ that are used in the next step (P2) where the noise parameters are sampled for.

\subsubsection{\label{sec:psd_model}Parametrization of the noise power spectral density}

In a complex measurement like LISA, it is safer to estimate the noise characteristics along with the signal, in order to avoid biases due to mismodeling of the noise PSD. Furthermore, we require some flexibility in the modeling, as the actual PSD may depart from the physical model described by Eq.~(\ref{eq:sim_psd_model}). To be as general as possible, we simply assume that the log-PSD is smooth enough to be modeled by cubic splines, borrowing from the \textsc{BayesLine} algorithm used to analyse LIGO science runs~\cite{Littenberg2014}. We found that writing the model for the log-PSD rather than for the PSD itself eases the estimation, because of the regularizing effect of the logarithm on the variance. While other more sophisticated models of the PSD can also be adopted in the context of Bayesian inference \cite{Edwards}, we leave the study of PSD modeling performance for future work. 
Let $\log f_j, \, \log S_j$ be the control points of the cubic spline. Then the model can be written $\forall j \in \left[ 1, \,  J \right]$ as:
\begin{eqnarray}
\label{eq:psd_model}
\log S_n(f) = \sum_{i=0}^{3} c_{i} \left( \log \frac{f}{f_j} \right)^{i} \text{ for } f \in \left[ f_{j} , f_{j+1} \right].
\end{eqnarray}
Although the number $N_s$ of control points can be a free parameter, in order to simplify our analysis and maintain a constant dimensionality, we fix them on a logarithmic grid lying in the interval $[10^{-n_0},\, 10^{-n_s} ]$ such that the grid spacing increases with the frequency:
\begin{eqnarray}
\log_{10} f_j = -n_0 + \frac{1-\alpha^j }{1-\alpha},
\end{eqnarray}
where $\alpha$ is a constant chosen such that $f_{J} = 10^{-n_s}$. Then, the PSD parameters to be estimated at step P2 are the control points $\boldsymbol{\theta}_{n} = \left( \log S_1 , \hdots , \log S_J\right)$, where typically $J = 30$. This is done via the Metropolis-Hastings method.

\subsection{\label{sec:mcmc_summary}Summary of the DA algorithm for UCB parameter estimation}

In previous sections we presented the general DA algorithm as well as the data analysis model that we use in the particular case of UCB parameter estimation. Here we summarize the main steps of the DA method for UCB. At iteration $i$, the process to follow is:
\begin{itemize}
\item[I] \textit{Missing data imputation.} For each gap, compute the covariance of neighboring observed points using Eq.~(\ref{eq:cov_from_psd}). Draw the data in the gap using their conditional distribution described in Eq.~(\ref{eq:miss_conditional}). 
Obtain a reconstructed time series $\boldsymbol{y}^{(i)}$.
\item[P] \textit{Sampling the parameter posterior distribution.}
\begin{itemize}
\item[P1] \textit{Sampling for PSD parameters.} Compute the DFT of the reconstructed time series $\boldsymbol{\tilde{y}}^{(i)}$ and form the model residuals $\boldsymbol{\tilde{y}}^{(i)}~-~\boldsymbol{h}(\boldsymbol{\theta}^{(i)}_{h})$. Sample the noise PSD parameters $\boldsymbol{\theta_{n}}$ using Whittle's likelihood in Eq.~(\ref{eq:whittle_like}) and the PSD model in Eq.~(\ref{eq:psd_model}) via a Metropolis-Hastings step and obtain the PSD update $S( \boldsymbol{\theta_{n}}^{(i)} )$.
\item[P2] \textit{Sampling for GW parameters}. Sample the GW parameters $\boldsymbol{\theta_{h}}$ in two steps: use a Metropolis-Hastings step with likelihood (\ref{eq:whittle_like}) to sample for intrinsic parameters, and sample for extrinsic parameters using Eq.~(\ref{eq:extrinsic_distrib}). Obtain $\boldsymbol{\theta_{h}}^{(i)}$. 
\end{itemize}
\end{itemize}
These steps are repeated until the distribution of intrinsic parameters reaches a stationary state.  

In order to implement this algorithm, some choices were made to increase its efficiency. We give some details below:

\paragraph{Size of conditional set.} The posterior step involves the inversion of the covariance matrix of the available observations. In the nearest neighbor approximation that we adopt, $N_g$ inversions are required, where $N_g$ is the number of gaps which ranges from 50 to 485 in our application. The choice of the size of the conditional set (which includes the two segments before and after the gap in our set-up) strongly affects the computational cost which scales as $N_{g} N_{c}^{3}$ where $N_{c}$ is the size of the conditional set. We have determined that for LISA noise PSD $N_{c} = 150$ is sufficient to have a faithful recovery of the spectrum, which corresponds to 25 minutes. This number is chosen to be conservative with respect to the decay time of the autocovariance function. For now, no parallelization nor optimization of the process have been done, leaving room for efficiency improvement in further studies.

\paragraph{Cadence of imputation steps.} In the DA algorithm presented in Eq.~(\ref{eq:da_algorithm}), the two steps usually do not have the same computational complexity. With the above choice of $N_{c}$, the imputation step requires a computation time two order of magnitude longer than the posterior step (a few seconds instead of a few tens of milliseconds), because it is dominated by the FFT computations. This may be cumbersome for large-scale MCMC algorithms. Although we would ideally like to perform an I-step after each P-step, we choose to tune their relative update cadence in order to decrease the computational burden. We find that performing an I-step every 100 P-steps is sufficient to obtain good posterior distributions in a couple of hours on a desktop computer.

\paragraph{MCMC set-up.} To sample the posterior distribution we use \textsc{ptemcee}, a parallel-tempered Markov chain Monte-Carlo sampler (PTMCMC) that dynamically adapts the temperature ladder \cite{Vousden2015}. It is built upon the \textsc{emcee} code \cite{Foreman-Mackey2012} which runs an ensemble of chains where the proposals for each chain is done based on the positions of other chains, using what is called a stretch move. We choose a uniform prior in $[0 ; \pi]$ and $[0 ; 2 \pi]$ for the colatitude $\theta$ and the longitude $\phi$ respectively. For the frequency, we choose a uniform but more restrictive prior centered on the amplitude maximum of the signal, with an interval range of about $2 \times 10^{-7}$ Hz. We use 10 different temperatures to reach a sufficient exploration of the posterior, and a number of chains equal to 4 times the number of dimensions (or 12), which is a minimal value to restrict the computation time.


\section{\label{sec:simulations}Numerical simulations}

In this section we describe the set of data and gap patterns that we use to assess the impact of missing data on parameter estimation, and introduce a way to tune the window function's smoothness in each case.

\subsection{Simulated signal and noise}
In the following we use a simulated one-year measurement of TDI channel $X$ sampled at 0.1 Hz, assuming an analytic Keplerian orbit for LISA's spacecrafts, and first-generation TDIs. We restrict the study to this single channel since we are interested in relative effects only. The gravitational signal is simulated for each phasemeter using the time-domain model described in Sec.~\ref{sec:sim_gw_model} and then recombined using Eq.~(\ref{eq:TDI1}). In order to concentrate on the effect of gaps, we assume constant and perfectly known arm lengths. 

We consider 3 compact, non-chirping galactic binary sources. We assume that these sources have the same sky localization, but different frequencies. We choose these frequencies to be sub-mHz, equal to 0.1, 0.2 and 0.5 mHz, as we shall see that the impact of gaps is only significant for lowest frequencies. Even if the probability to observe such systems with sufficient SNR is not high, we consider this case for the sake of assessing the qualitative impact of gaps on parameter estimation. The amplitude of the sources are chosen so that their SNR is about 46 in the X channel alone. Note that for such low frequencies, chirping can be neglected. Therefore we do not include the frequency derivative $\dot{f}_0$ in the set of parameters to estimate. Sky location is chosen arbitrarily, as we do not expect source localization to have a major influence on the general results of this study given that the simulation time spans an entire orbit. Table~\ref{tab:source_parameters} summarizes the values of the source parameters.

 \begin{table}[ht]
\caption{\label{tab:source_parameters}Values of the UCB source parameters used in the simulations. While any other parameter remains the same, the sources in the 3 data sets differ from their amplitudes (first row) and frequencies (second row). All sources have the same SNR of about 46.} 
 \begin{ruledtabular}
 \begin{tabular}{l c}
Parameter & Value \\
\hline
Amplitude [strain $\times 10^{-20}$] & 15, 2.0, 0.2  \\ 
Frequency [mHz] & 0.1, 0.2, 0.5 \\
Ecliptic latitude [rad] & 0.47 \\
Ecliptic longitude [rad] & 4.19 \\
Inclination [rad] & 0.179 \\
Initial phase [rad] & 5.78 \\
Polarization [rad] & 3.97 \\
 \end{tabular}
 \end{ruledtabular}
 \end{table}

\subsection{Simulated gap patterns}

We consider two gap scenarios in the following. The first one is a planned interruption schedule, simulating generic maintenance periods such as antenna repointing (see Sec.~\ref{sec:intro}). 
While we do not have yet precise information about what the future maintenance cycle will be, we conservatively assume rather frequent periodic interruptions: we simulate one interruption every 5 days, lasting approximately 1 hour. Given that some flexibility will be possible on the gap times, and that all operations may not last the same time, we allow both the gap time locations and their duration to randomly vary. The gap start times follow a periodic pattern with deviations modeled by a Gaussian distribution with a standard deviation of 1 day. The duration is also a Gaussian distribution with mean 1 hour and standard deviation 10 minutes. 

The second gap pattern models unplanned interruptions, due to any glitch events preventing the instrument from properly acquiring the measurement.
Based on LISA Pathfinder feedback, these kind of events are likely to occur at an average rate of 0.78/day~\cite{Armano2018a}. To be conservative, we assume 1 event per day. The number of events in a given interval is assumed to follow a Poisson distribution, so that the intervals between gaps follow an exponential distribution. We assume that each gap lasts about 10 minutes, with a standard deviation of 1 minute. In the following we label the two gap patterns ``five-day periodic gaps" and ``daily random gaps" respectively, and we summarize their characteristics in Table~\ref{tab:gap_patterns}. 

 \begin{table}[ht]
 \caption{\label{tab:gap_patterns} Two types of gap patterns are considered: one models planned events such as antenna operation gaps, while the second one models unplanned events such as glitch masking.}
 \begin{ruledtabular}
 \begin{tabular}{l c c }
 &  Five-day periodic gaps & Daily random gaps \\
\hline
Occurrence & 5 days $\pm$ 1 day & 1 day  $\pm$ 1 hour\\
Duration & 1 hour $\pm$ 10 min & 10 min $\pm$ 1 min\\
Loss fraction &  0.8\% &  0.7\%
 \end{tabular}
 \end{ruledtabular}
 \end{table}

\begin{figure}[ht]
 \includegraphics[width=0.5\textwidth, trim={0.3cm 0cm 0cm 0cm},clip]{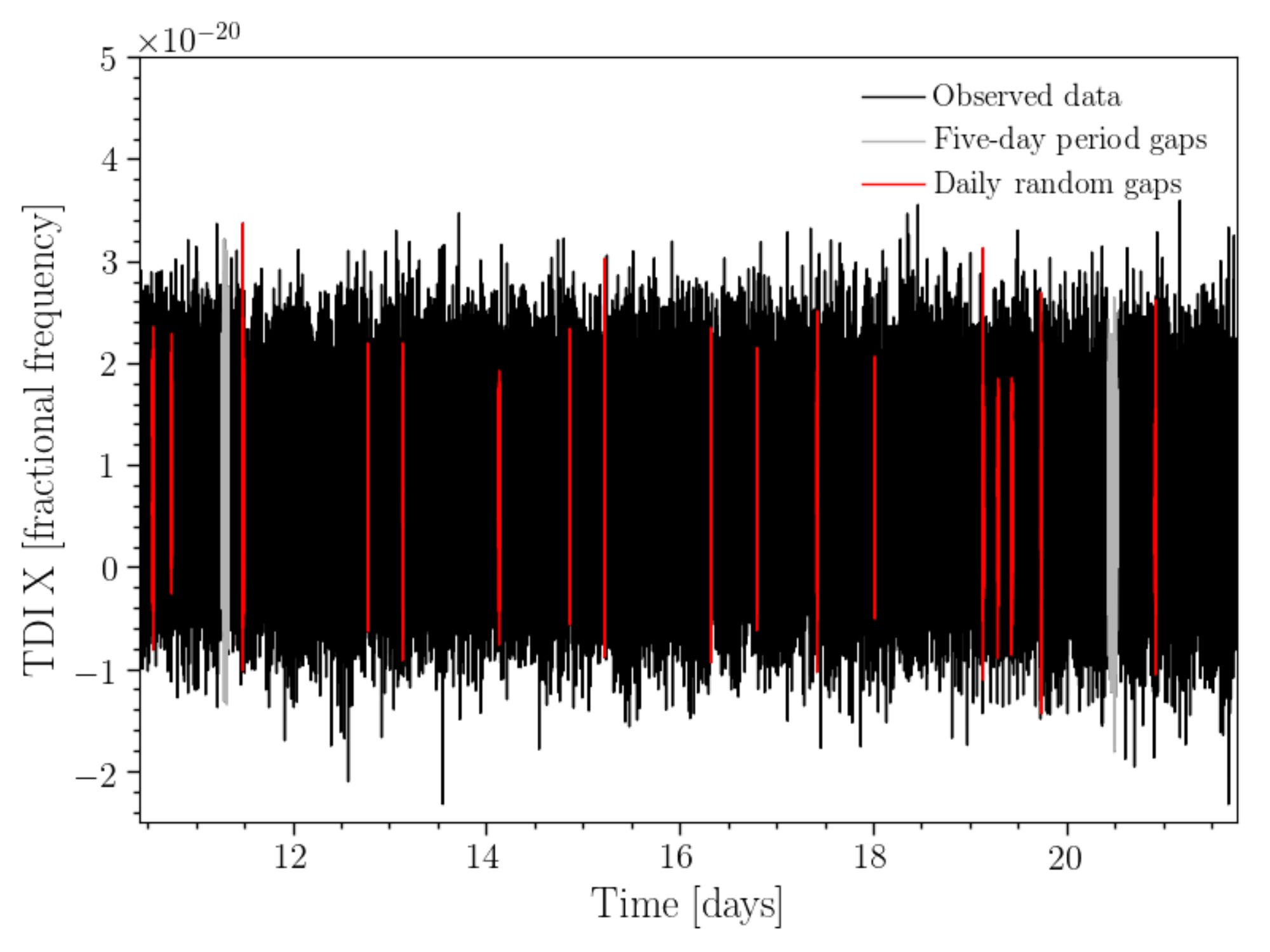}%
  \caption{\label{fig:time_series}Segment of a simulated times series with observed data in black, missing data in gray for five-day periodic gaps and in red for Daily random gaps. In scenario A, gaps are 5 times longer and 5 times less frequent than in scenario B.}      
 \end{figure}

It is worth noting that the two gap patterns have almost the same loss fraction (less than 1\%) but strong differences in gap occurrences and duration.
A visual insight is provided in Fig.\ref{fig:time_series} where we plot an extract of a simulated data representing TDI X amplitude as a function of time, expressed in fractional frequency. Data lying inside gaps are plotted in gray for five-day periodic gaps and in red for daily random gaps. The remaining observations are shown in black. This plot highlights the difference of gap occurrences in the two patterns.

\subsection{\label{sec:windowing_optimization}Optimization of windowing}

As mentioned in Sec.~\ref{sec:windowing}, the impact of gaps can be mitigated using a window function smoothly decaying at the gap edges. Hence we have to choose the amount of smoothness. For a given source, a given noise and a given gap pattern, it is actually possible to find an optimal value. In this section we present a way to perform such an optimization and adopt it in the simulations as our baseline to assess the impact of gaps. 

\begin{figure}[ht]
 \centering
 \begin{tabular}{c}
  \hspace{0.8cm} Five-day periodic gaps \\
  \includegraphics[width=0.49\textwidth]{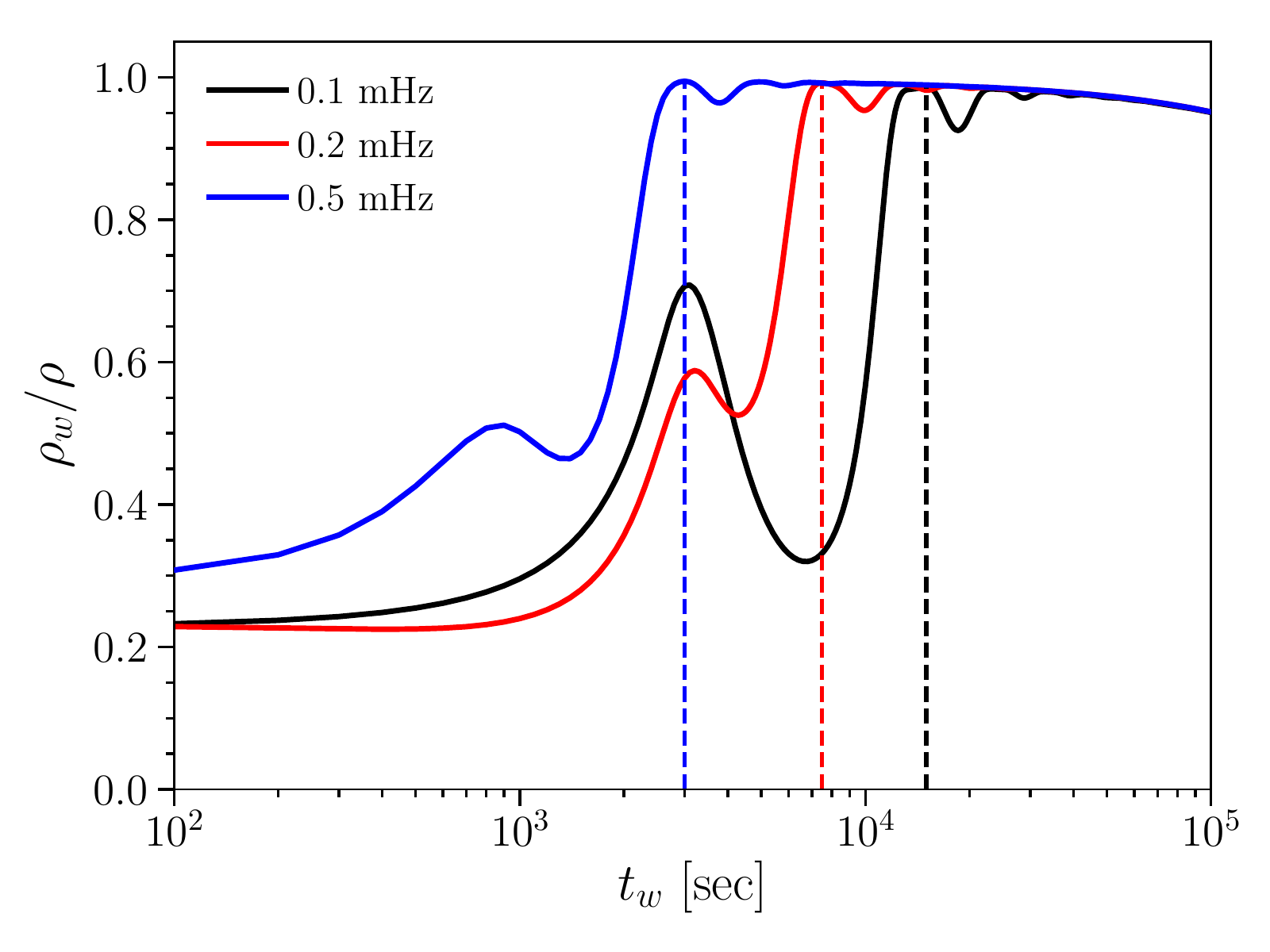} \\
  \hspace{0.8cm} Daily random gaps \\
 \includegraphics[width=0.49\textwidth]{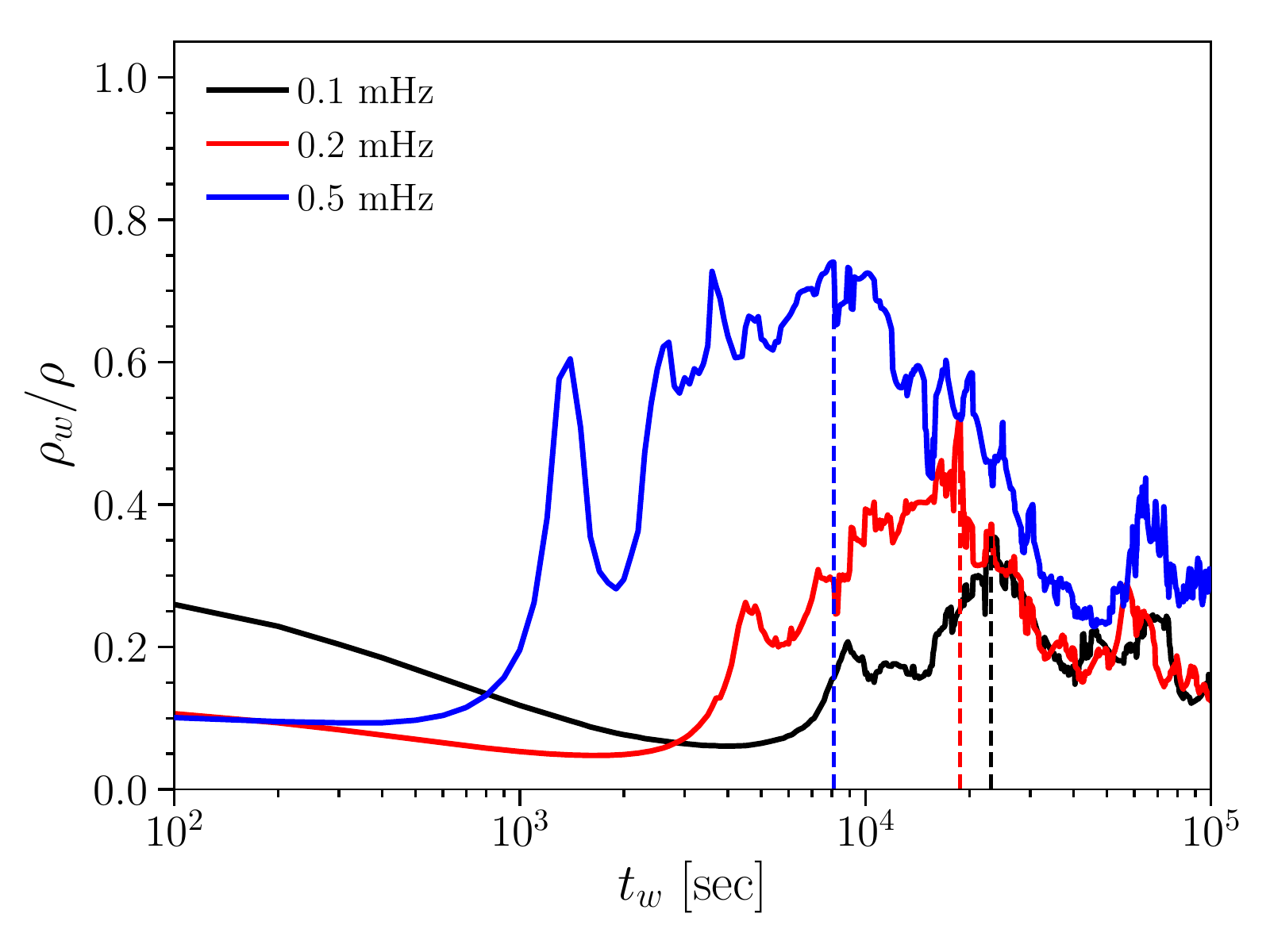}
 \end{tabular}
 \caption{\label{fig:t_w}Effective SNR as a function of the smoothing time $t_w$ calculated with Eq.~(\ref{eq:SNR_eff}) in the case of Five-day periodic gaps (top panel) and B (bottom panel), normalized by the optimal SNR value obtained for complete data. The calculation is done for 3 sources of frequency 0.1 mHz (black), 0.2 mHz (red) and 0.5 mHz (blue). The dashed vertical lines show the value of $t_w$ where the maximum is reached.}
 \end{figure}

We use a Tukey-like window, such that each segment of available data of length $T_s$ is tappered with the a window function parametrized by the smoothing time $t_w$:
\begin{equation}
w_{{T_s}}(t) \equiv 
\begin{cases} 
      \frac{1}{2}\left[ 1 - \cos\left(2\pi \frac{t}{2t_w}\right) \right]  & 0 \leq t < t_w  \\
      1 & t_w \leq  t <  T_s - t_w \\
      \frac{1}{2} \left[ 1 - \cos\left( 2 \pi  \frac{t-T_s + 2 t_w}{ 2 t_w} \right) \right] & T_s - t_w \leq t < T_s \\
      0 & \text{otherwise},
   \end{cases}
\end{equation}
such that the full window function is 
\begin{eqnarray}
w(t) = \sum_{s=1}^{N_s} w_{{T_s}}\left( t - t_{s} \right),
\end{eqnarray}
where $t_s$ is the starting time of segment $s$ (i.e. the end of the previous gap). 
In order to choose the optimal smoothing time $t_w$ (i.e. the time controlling the transition length between 0 to 1 and conversely), we can resort to the effective SNR defined in Eq.~(\ref{eq:SNR_eff}), and plot it as a function of $t_w$. We find that there is a value $t_{w,\mathrm{opt}}$ that maximizes the effective SNR.
Choosing $t_w = t_{w,\mathrm{opt}}$ ensures a trade-off between the minimization of noise leakage and the limitation of SNR loss due to tapering. 
In Fig.~\ref{fig:t_w} we plot the effective SNR for the 3 considered sources, both for five-day periodic gaps (top) and B (bottom). In order to better assess the impact of gaps, the SNR is normalized by the complete data SNR.

These plots show that the function $\rho_{w}(t_w)$ first starts to increase (however not always monotonically), reaches a maximum and then slowly decreases as $t_w$ continues to grow. This behavior can be easily understood: for values close to $t_w = 0$ the window function is rectangular and has rough edges, generating a large leakage effect, thus increasing the denominator in Eq.~(\ref{eq:SNR_eff}). For values larger than the optimal threshold $t_{w,\mathrm{opt}}$ (indicated by dashed vertical lines on the figure), further smoothing the window does not better cancel the leakage while making the tapered signal loose a bit of its power (i.e. the numerator in  Eq.~(\ref{eq:SNR_eff}) is then decreasing). In addition, we see that the amount of leakage is larger for Daily random gaps (i.e. when the gaps are shorter and more frequent) than for five-day periodic gaps: while for periodic gaps the value of the effective SNR at $t_{w,\mathrm{opt}}$ is close to 100\% of the complete-data SNR, for random gaps it drops to about 70\% for $f_0 = 0.5$ mHz and to about 30\% for $f_0 = 0.1$ mHz. Comparing the 3 curves also indicates that the lower the frequency, the more the SNR is affected by leakage.


\section{\label{sec:estimation_results}Estimation results}

After simulating the 3 sources and the 2 gap patterns as described in Sec.~\ref{sec:simulations}, we aim at recovering the signal and noise parameters from these simulations. This is done by sampling their posterior distribution using the PTMCMC algorithm outlined in Sec.~\ref{sec:mcmc_summary}. In this section we present the results that we obtain, in the case of complete data and gapped data, using the windowing and the DA method. 

\subsection{\label{sec:parameter_results}Results of parameter estimation for one single source}

\begin{figure*}[ht]
 \centering
 \begin{tabular}{ c | c }
 \multicolumn{2}{c}{
  \includegraphics[width=0.7\textwidth, trim={0cm 0cm 0cm 0cm},clip]{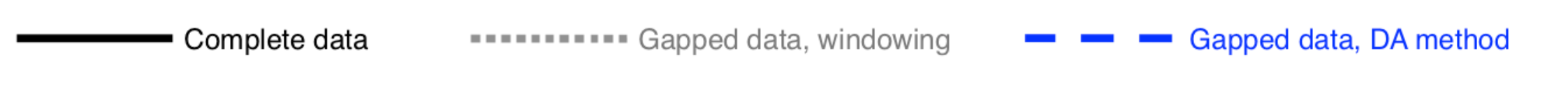}
  } \\
 $f_0 = 0.1$ mHz, Five-day periodic gaps & $f_0 = 0.1$ mHz, Daily random gaps \\
 \includegraphics[width=0.25\textwidth, trim={0cm 0cm 0cm 0cm},clip]{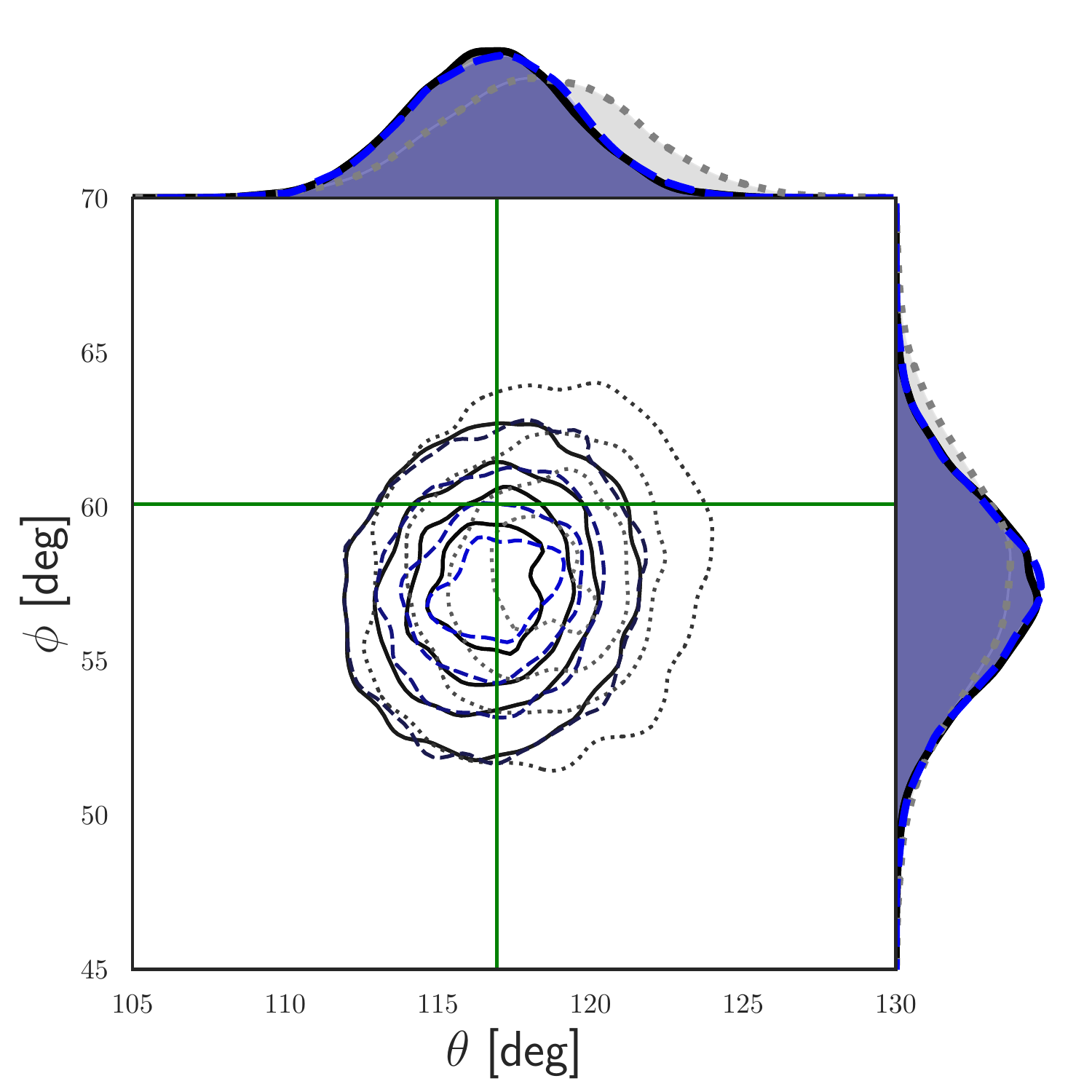}
\hfill
 \includegraphics[width=0.225\textwidth, trim={0cm 0.5cm 0cm 0cm},clip]{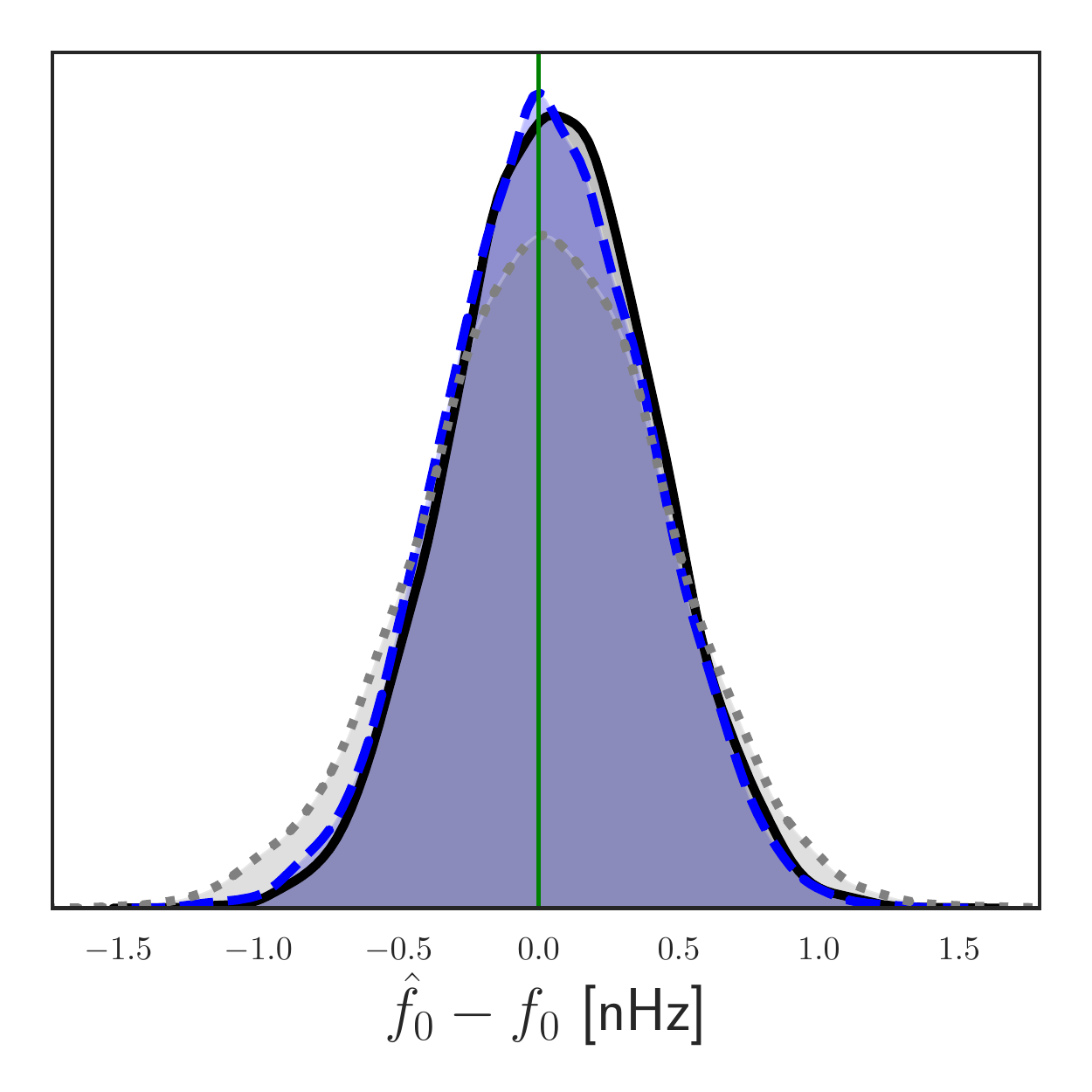}
&
  \includegraphics[width=0.25\textwidth, trim={0cm 0cm 0cm 0cm},clip]{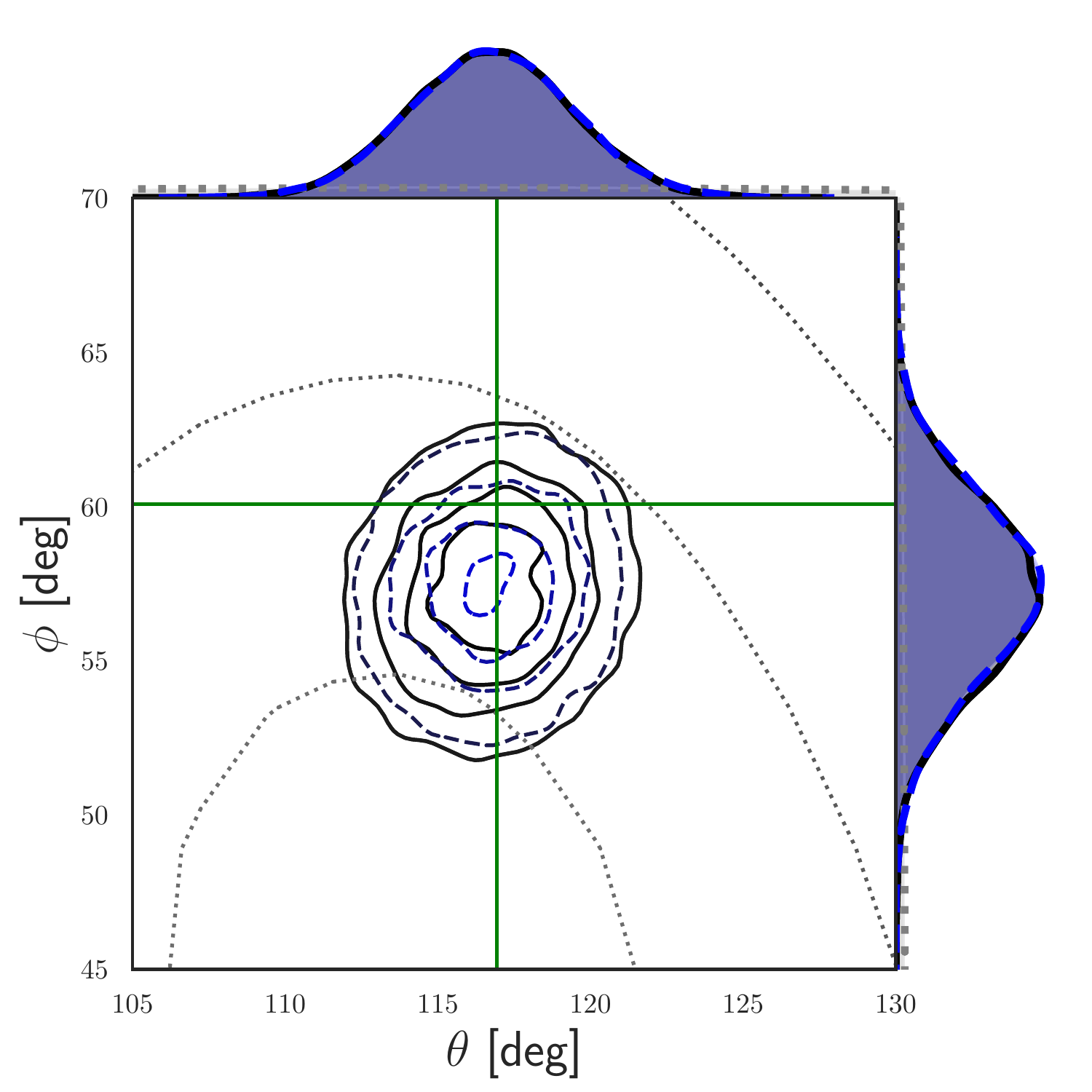}
\hfill
 \includegraphics[width=0.225\textwidth, trim={0cm 0.5cm 0cm 0cm},clip]{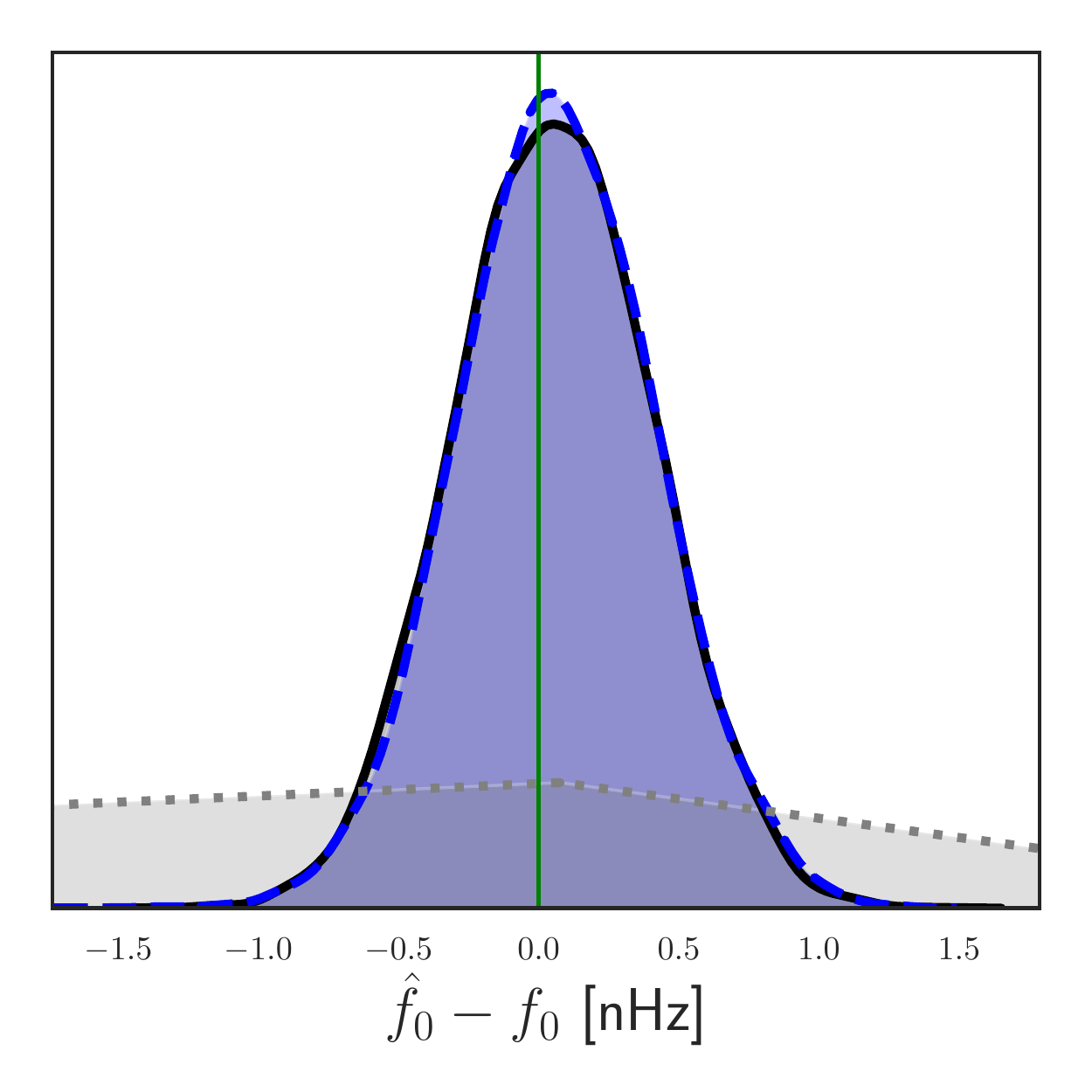}
 \\ \hline
$f_0 = 0.2$ mHz, Five-day periodic gaps & $f_0 = 0.2$ mHz, Daily random gaps
\\ 
\includegraphics[width=0.25\textwidth, trim={0cm 0cm 0cm 0cm},clip]{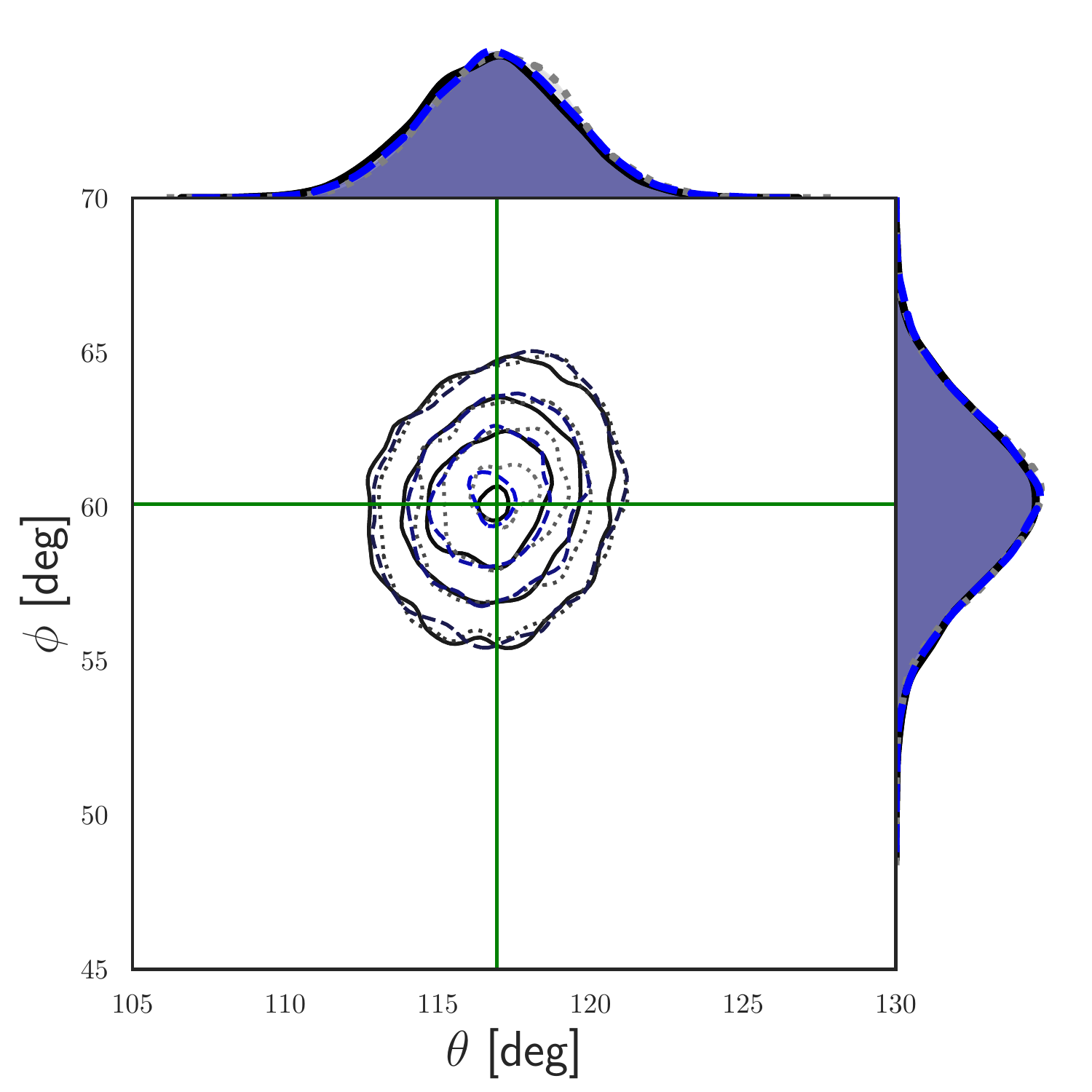}
\hfill
\includegraphics[width=0.225\textwidth, trim={0cm 0.5cm 0cm 0cm},clip]{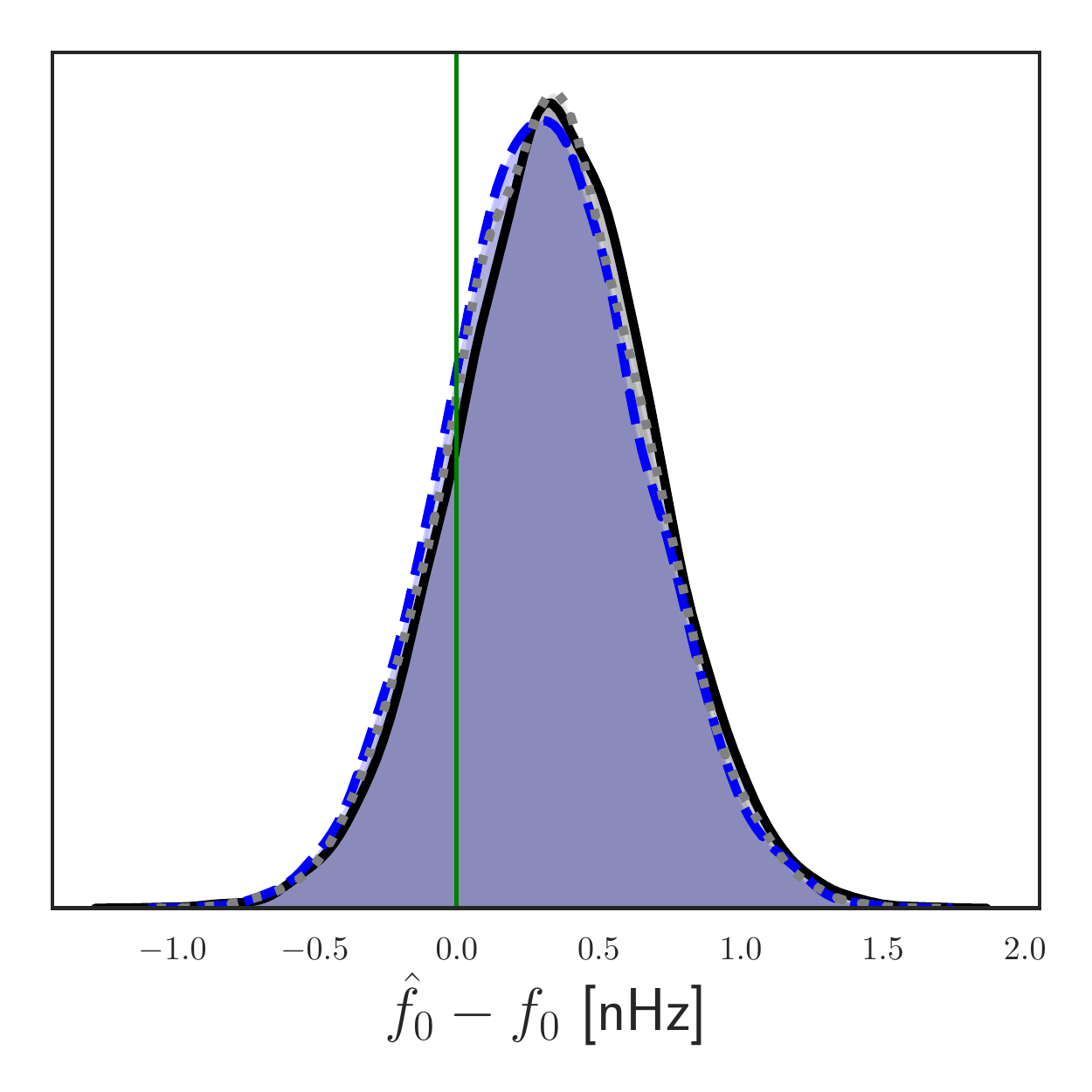}
&
\includegraphics[width=0.25\textwidth, trim={0cm 0cm 0cm 0cm},clip]{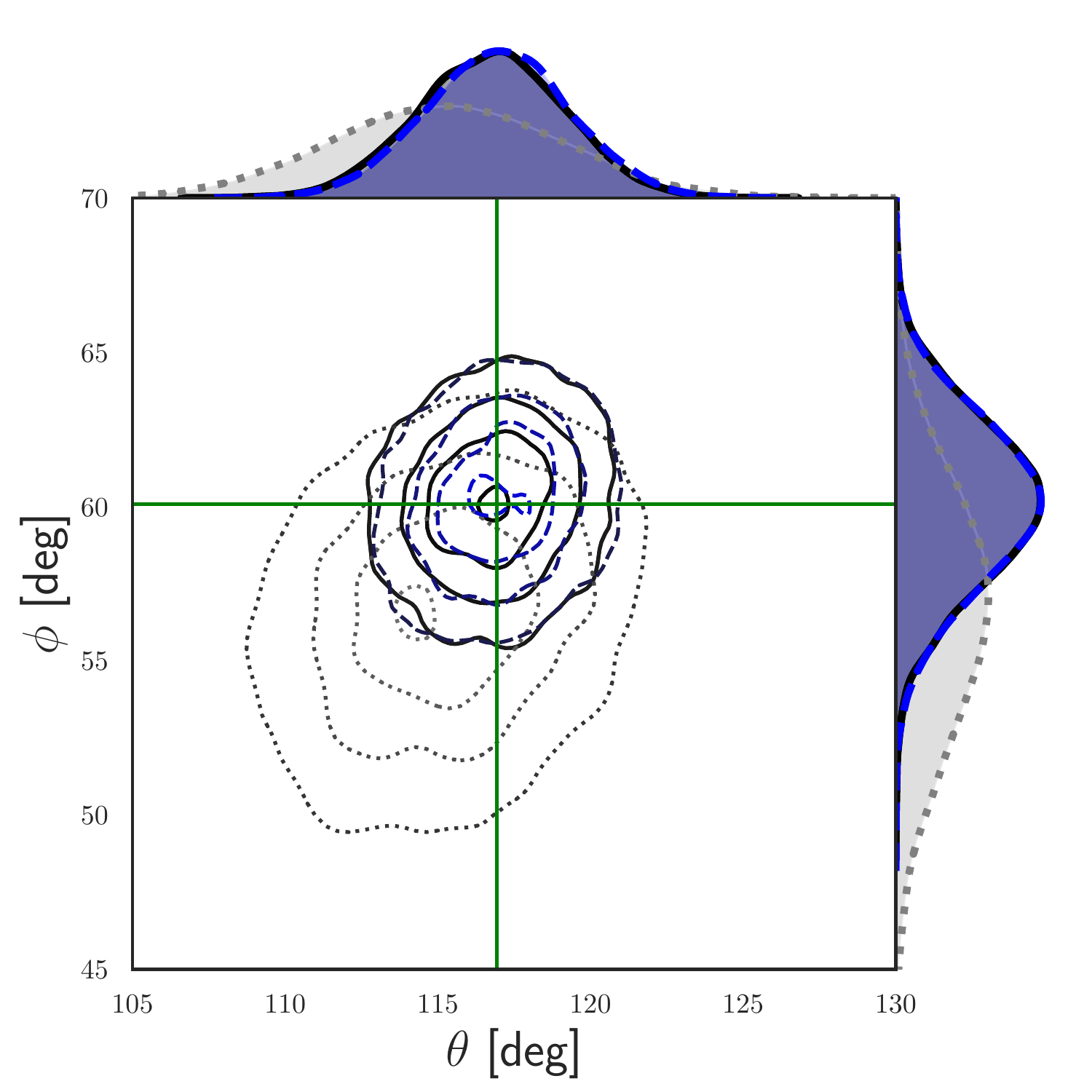}
 \hfill
\includegraphics[width=0.225\textwidth, trim={0cm 0.5cm 0cm 0cm},clip]{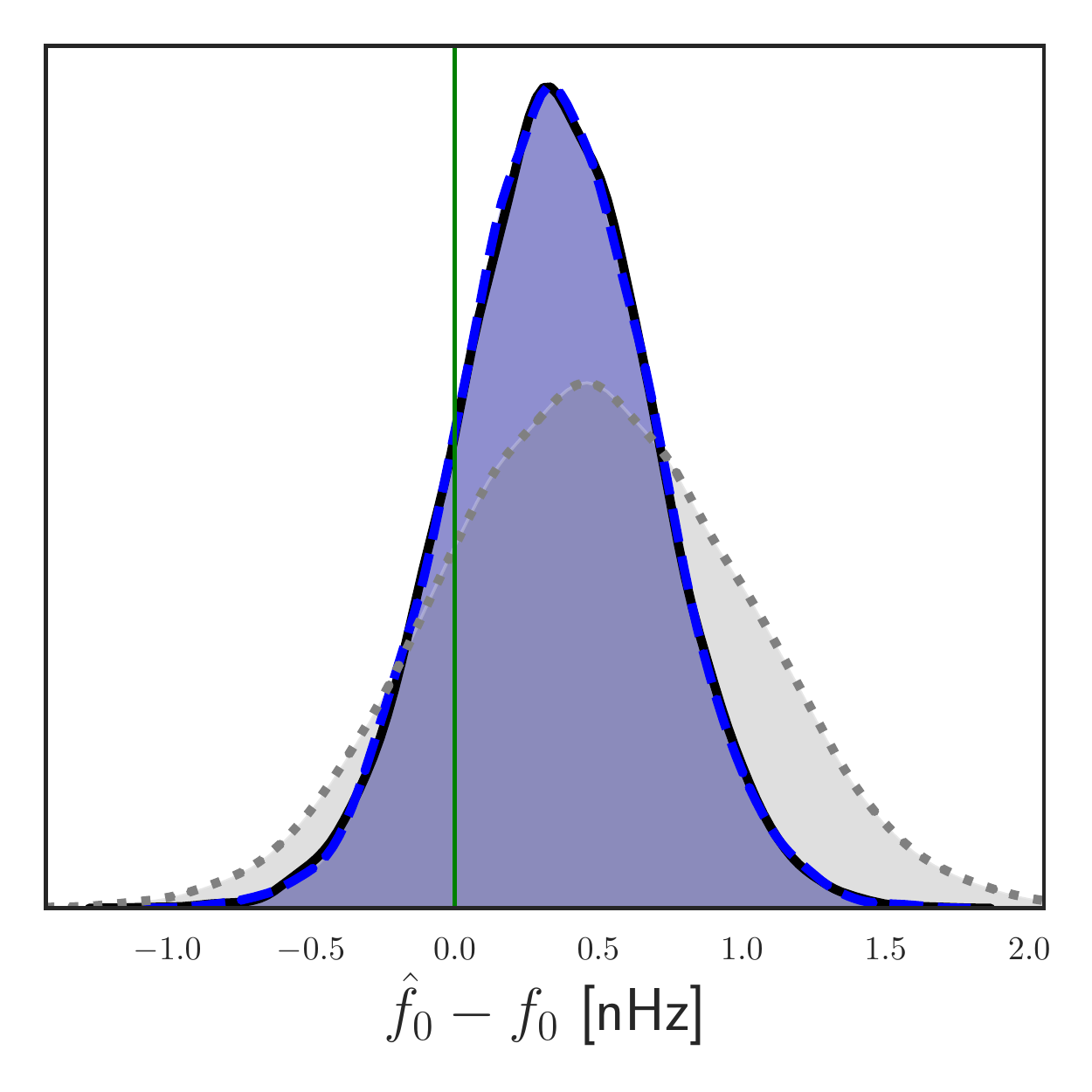}
\\ \hline
$f_0 = 0.5$ mHz, Five-day periodic gaps & $f_0 = 0.5$ mHz, Daily random gaps
\\
\includegraphics[width=0.25\textwidth, trim={0cm 0cm 0cm 0cm},clip]{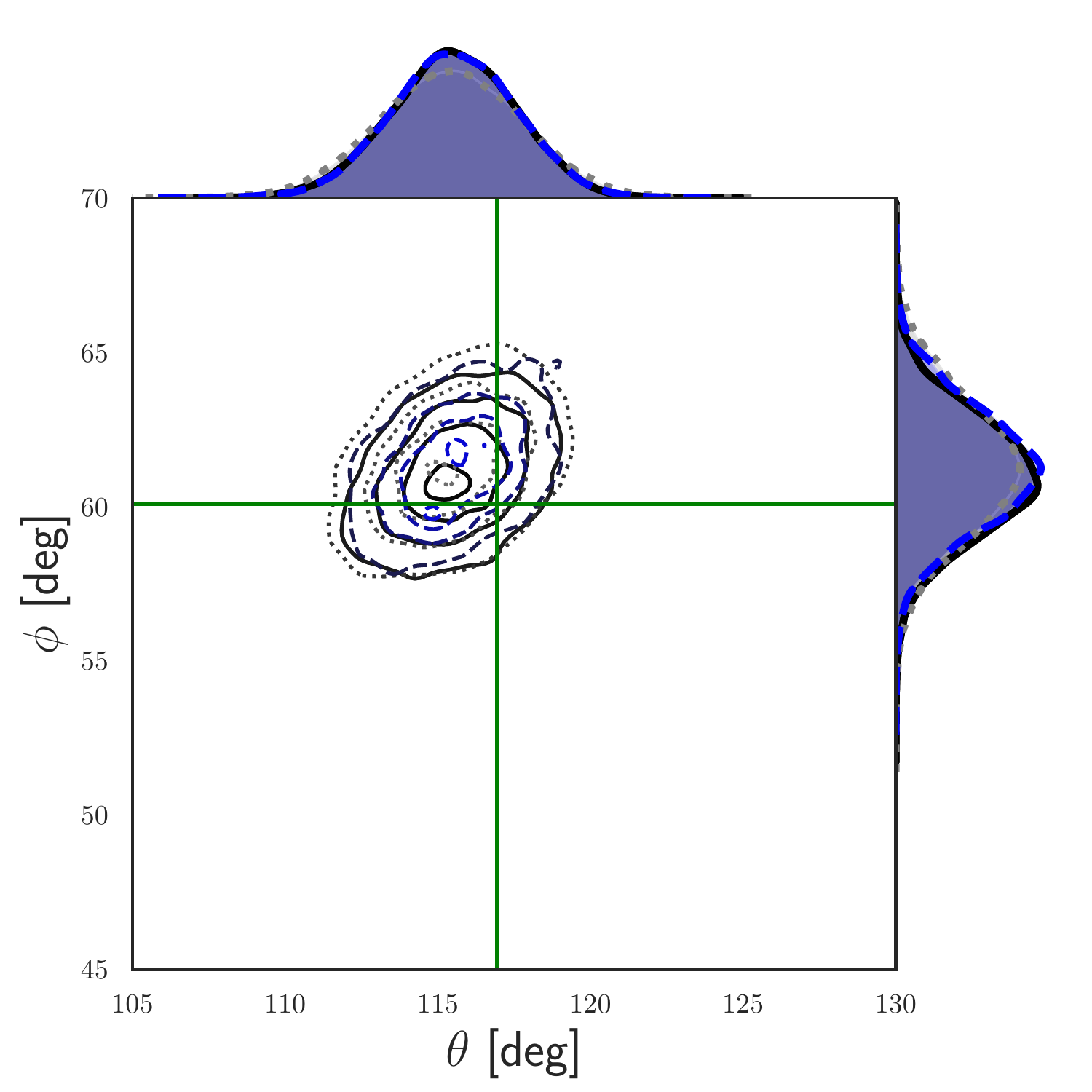}
\hfill
\includegraphics[width=0.225\textwidth, trim={0cm 0.5cm 0cm 0cm},clip]{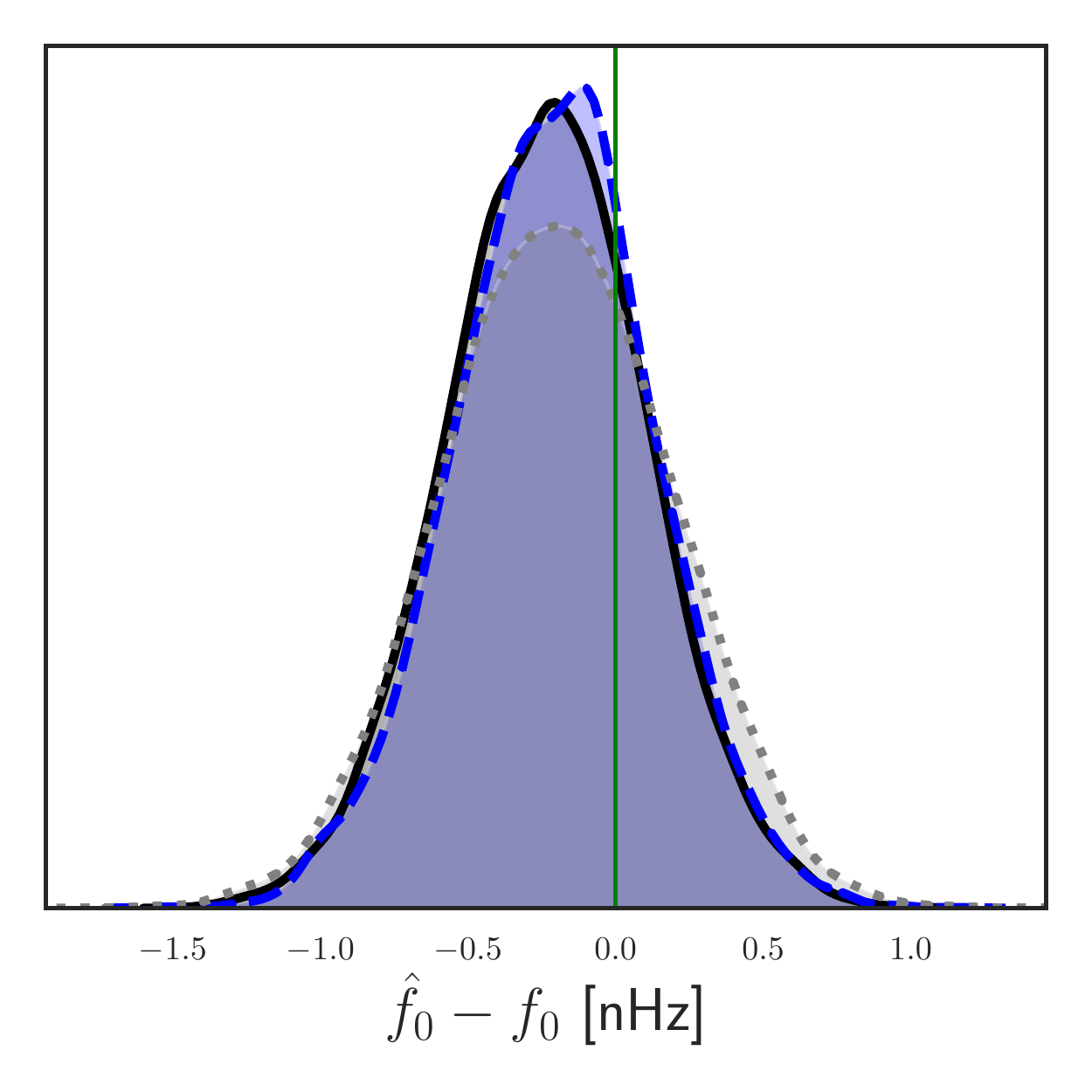}
&
\includegraphics[width=0.25\textwidth, trim={0cm 0cm 0cm 0cm},clip]{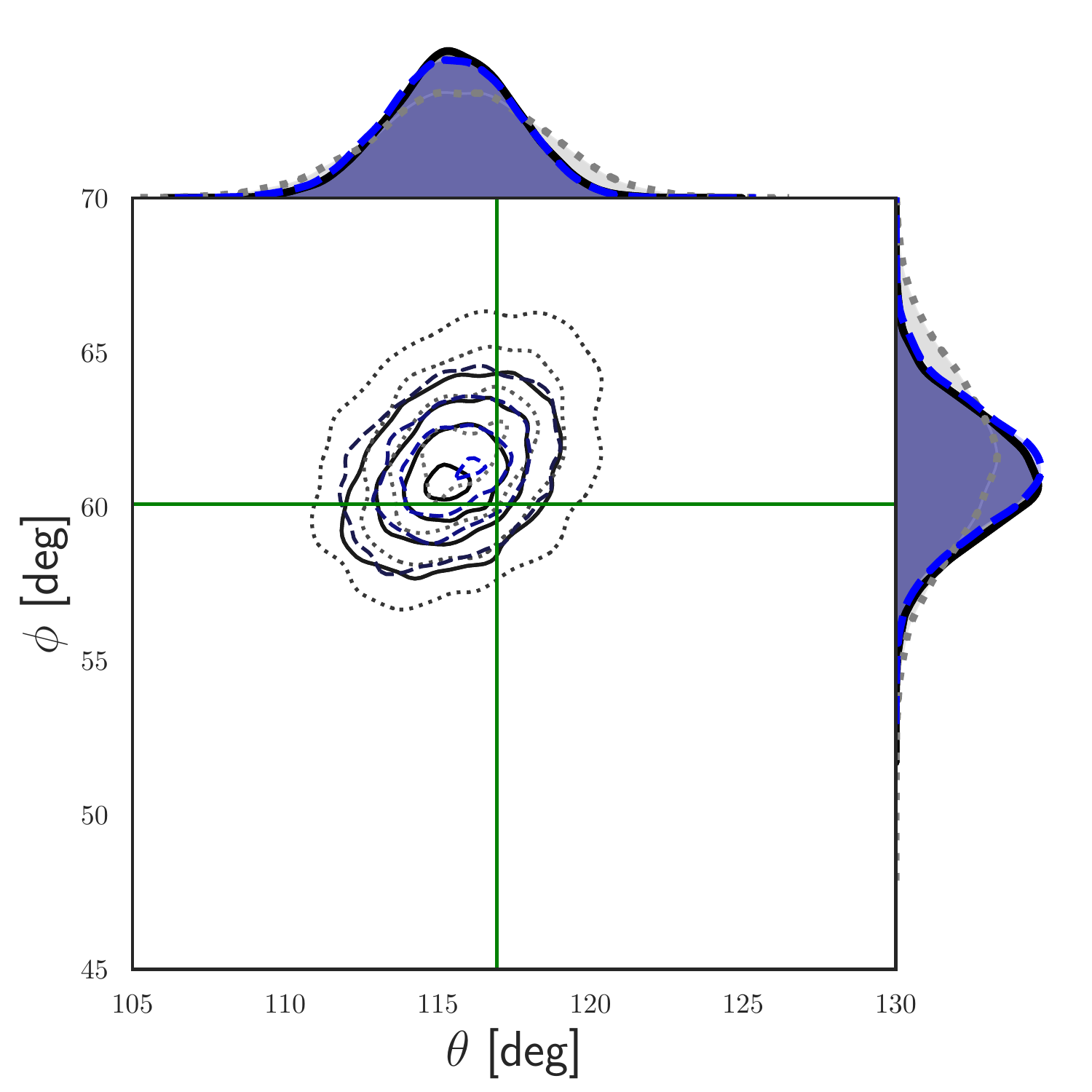}
\hfill
\includegraphics[width=0.225\textwidth, trim={0cm 0.5cm 0cm 0cm},clip]{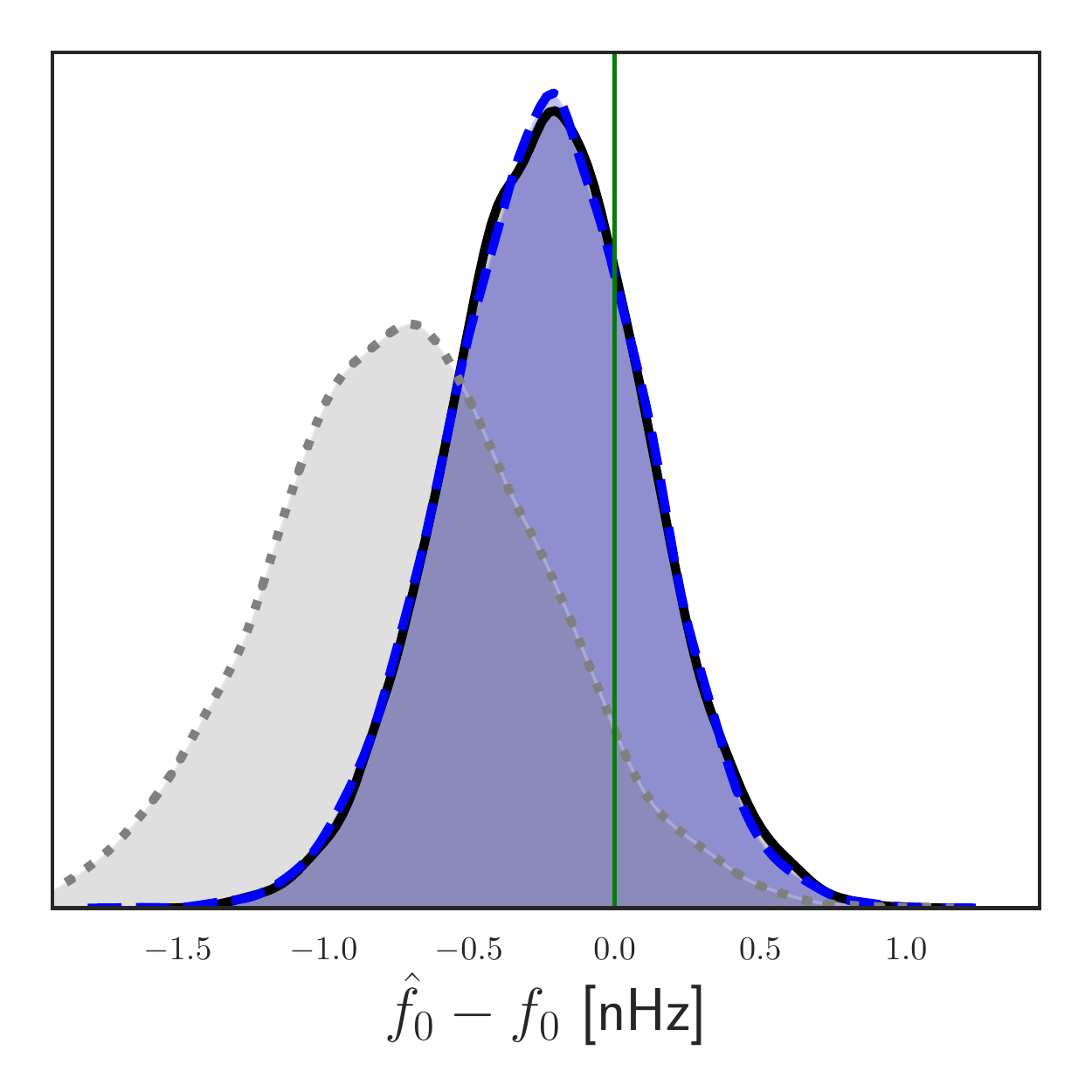}
\\
\end{tabular}
 \caption{\label{fig:histograms}Result of PTMCMC sampling of the posterior distribution for ecliptic colatitude $\theta$, ecliptic latitude $\phi$, and GW frequency $f_0$ for UCB sources with frequencies $f_0 = 0.1$ mHz (first row), $f_0 = 0.2$ mHz (second row), and $f_0 = 0.5$ mHz (third row). The left-hand side panels correspond to periodic gaps and the right-hand side panels to random gaps. The estimated posterior distributions obtained from full data are shown by solid black curves. The ones obtained from gapped data are shown in dotted gray when using the windowing method, and in dashed blue when using the DA method.}
 \end{figure*}

We first consider the case where one single source is present in each data series. For each source, a first estimation is done by running the PTMCMC algorithm on complete data. This provides the ``best case" baseline to compare the results. Then we introduce gaps in the data (for each pattern), and we perform a second estimation by running the PTMCMC algorithm using the windowed data, where the smoothing time $t_w$ is chosen equal to its optimal value. A third estimation is done using the data augmentation method presented in Sec.~\ref{sec:da}. We plot the results of the three estimations in Fig.~\ref{fig:histograms}, as histograms representing the joint posterior distribution of ($\theta$, $\phi$) and the posterior distribution of $f_0$.



Starting with $f_0 = 0.1$ mHz (first row), the windowing method (gray dotted line) applied to five-day periodic gaps (right-hand side panel) yields a slightly larger posterior distribution than the complete data series (solid  black), suggesting a slight effect of noise power leakage. The DA method (dashed blue) gives results comparable to the complete data case, with a similar variance. If we now consider daily random gaps (right-hand side panel), the difference between the two methods is more obvious. In this case, the parameters cannot be recovered using the windowing method. The noise power leakage is too large for the signal to stand out, and the obtained posterior distribution is dominated by noise. However, the DA method yields a posterior distribution that is consistent with the injection, and similar to the case without any gaps. This quasi-similarity is expected for data losses lower than 1\%. 

Now considering $f_0 = 0.2$ mHz (second row), in the case of five-day periodic gaps the difference between posterior distributions obtained with the two methods is hardly visible, and they are similar to the case of complete data. However, for daily random gaps, while the windowing method is still able to recover the injection, we observe a broadening of the distribution with standard deviations increasing by 50\% for all parameters. This is due to the leakage effect that is not completely removed as shown by the bottom panel of Fig.~\ref{fig:t_w}, which also indicates a loss of SNR by about the same amount. Applying the DA method allows us to obtain a posterior that is close to the case of complete data.

For $f_0 = 0.5$ mHz, the differences between complete data and gapped data are getting narrower, as well as the differences between the results of the two methods. Again, this observation can be related to Fig.~\ref{fig:t_w} where we saw that for the highest frequency the impact of leakage becomes insignificant.

\subsection{Results of PSD estimation}

 \begin{figure*}[ht]
 \begin{tabular}{ c c }
\hspace{0.8cm} Five-day periodic gaps &\hspace{0.8cm} Daily random gaps \\
 \includegraphics[width=0.49\textwidth,  trim={0cm 0.5cm 0cm 0cm},clip]{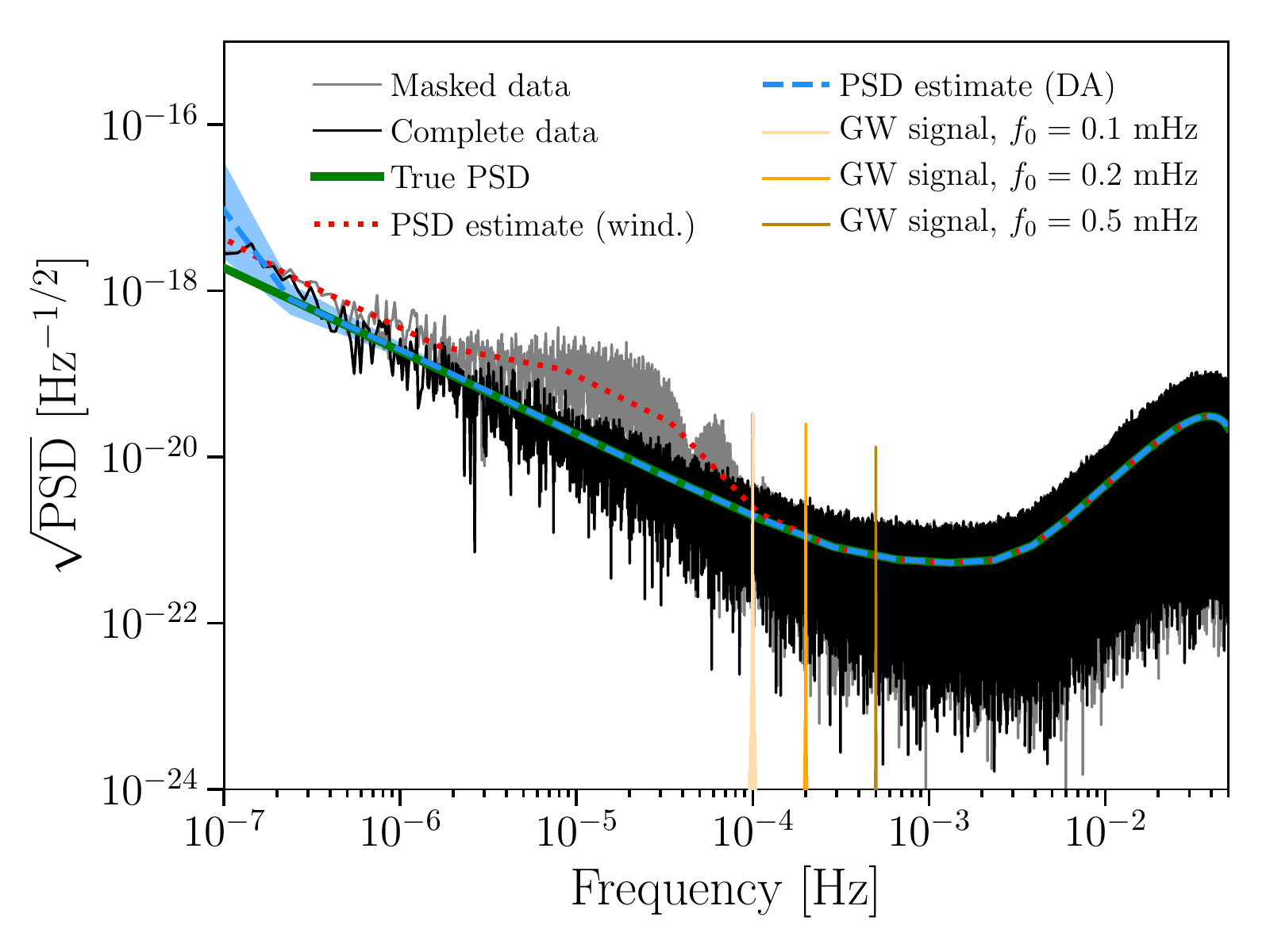}%
 &
 \includegraphics[width=0.49\textwidth,  trim={0cm 0.4cm 0cm 0cm},clip]{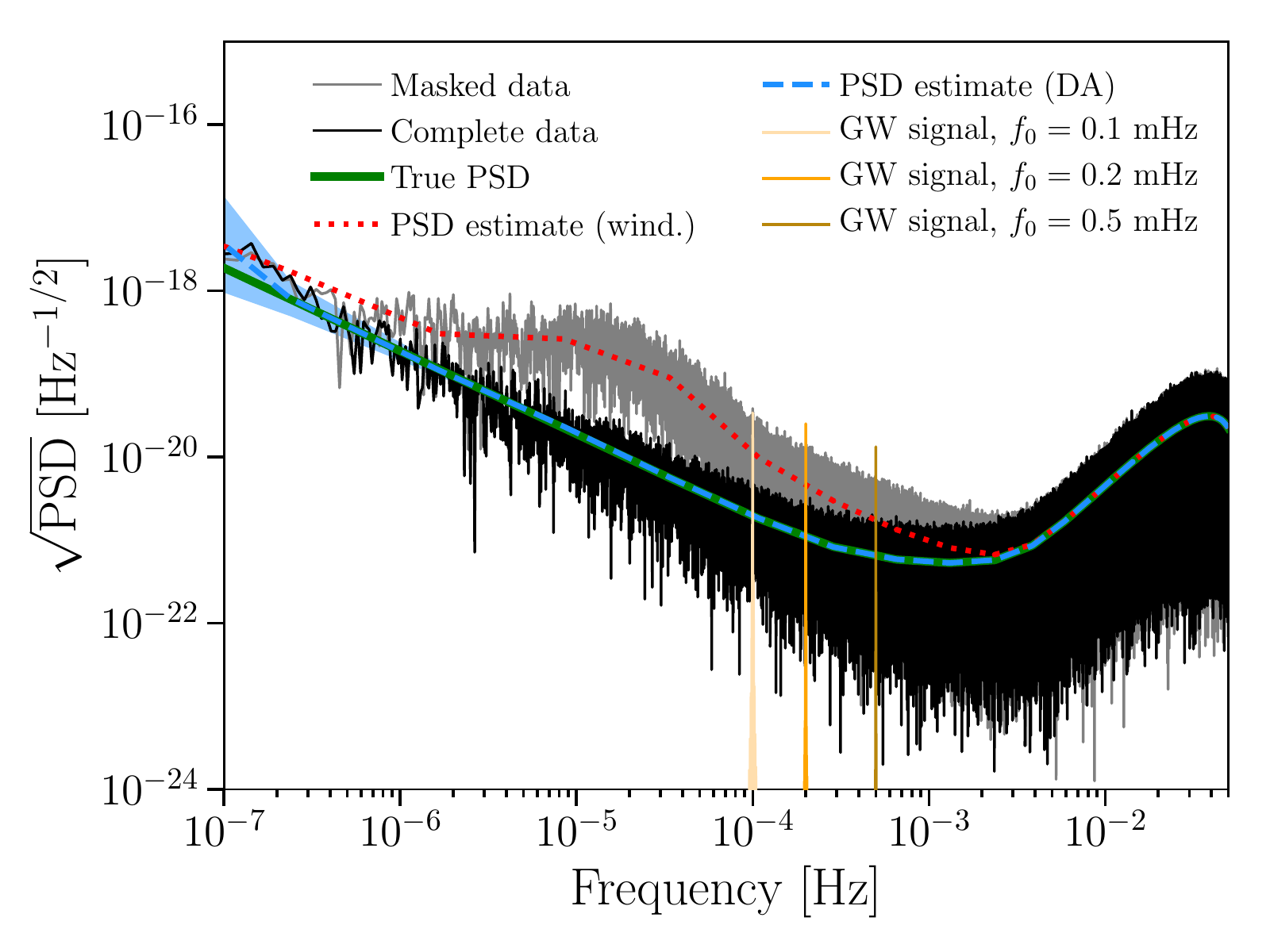}
  \end{tabular}
 \caption{\label{fig:psd_estimation}Results of PSD estimations with gapped data, with five-day periodic gaps (left-hand side) and daily random gaps (right-hand side). Dotted red curves show PSD estimates obtained with the windowing method, and dashed blue curves the ones obtained with the DA method. They are compared to the true PSD represented by solid green curves. The window method estimates are affected by leakage effects due to the gapped observation window, while the DA method yields an unbiased estimates in most of the frequency band. Black and gray solid lines respectively represent periodograms of complete data and periodograms of gapped data. The peaked curves in orange shades correspond to GW sources at 0.1 mHz, 0.2 mHz and 0.5 mHz. For a 1-year integration time, their signal stand out of the noise with five-day periodic gaps, but is overwhelmed by noise leakage with daily random gaps for $f_0 = 0.1$ mHz.}
 \end{figure*}

In this section we analyze the results of the noise PSD estimation that is obtained with the model described in Sec.~\ref{sec:psd_model}. The PSD parameters are estimated from the model residuals at each posterior step of the Gibbs algorithm. The estimated PSD posteriors are presented in Fig.~\ref{fig:psd_estimation}. 

In this figure we plot the periodograms of the completed data (black) and the gap-windowed data (gray), along with the estimated PSD using the windowing method (dotted brown), and using the DA method (dashed blue). In both cases the estimates are obtained by using the maximum a posteriori (MAP) estimator. The $3\sigma$ confidence intervals (light blue areas) are computed using the sample variance of the posterior distribution. For comparison, the true PSD used to generate the data (see Eq.~(\ref{eq:sim_psd_model})) is shown in solid green. 

\begin{figure}
 \includegraphics[width=0.5\textwidth, trim={0cm 0.4cm 0cm 0cm}, clip]{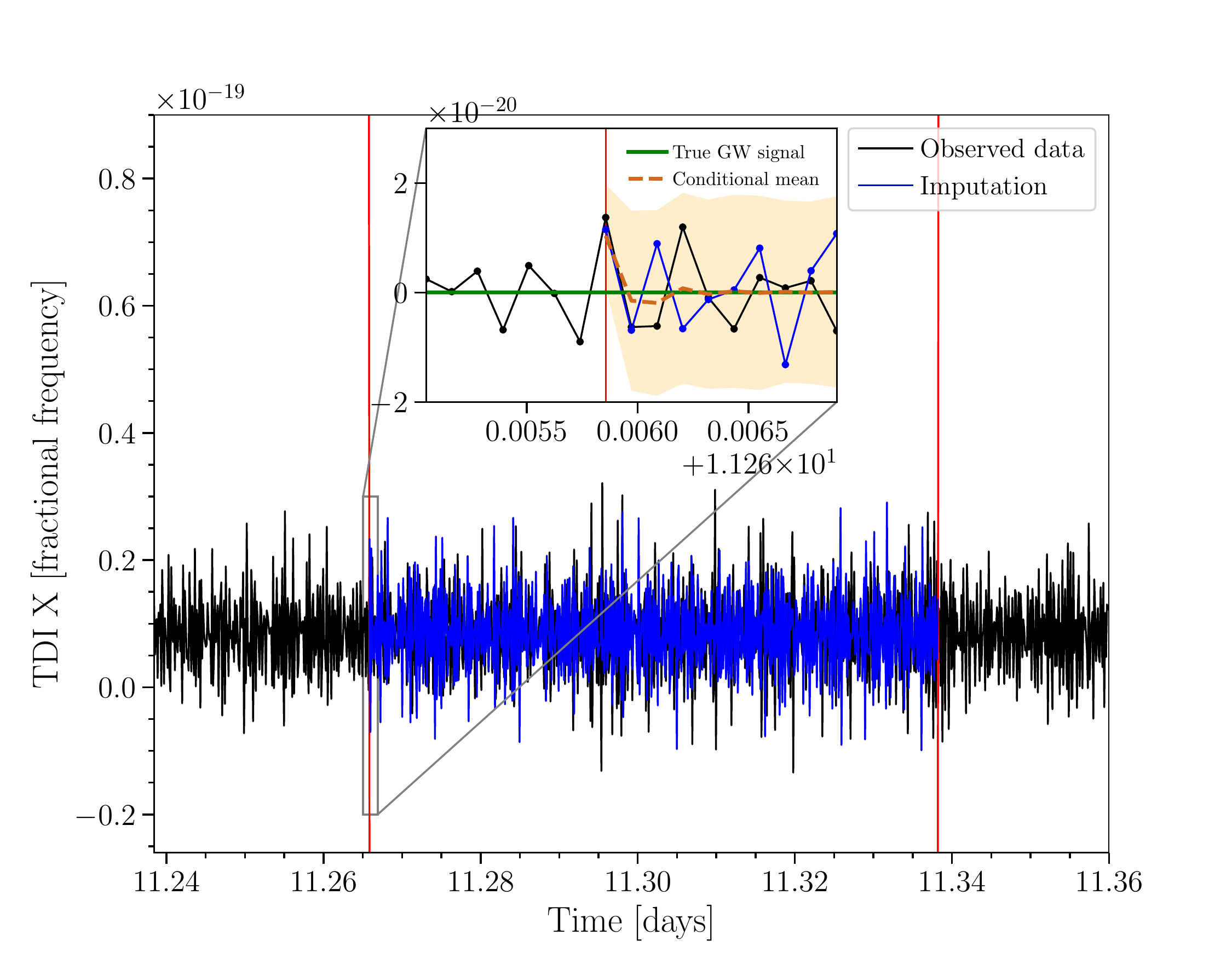}%
 \caption{\label{fig:time_imputation}Results of one gap imputation draw in the time domain after the MCMC chains have reach stationarity. This draw is obtained for a periodic gap pattern and a source frequency $f_0~=~0.1$ mHz. The noise statistics are preserved inside the gap, allowing us to accurately sample the PSD when Fourier-transforming the imputed data. The zoomed inset shows the GW signal (green) and the estimated conditional mean $\mu_{m | o}$ inside the gap (dashed orange), taking into account both noise correlations and deterministic signal. The colored area represents the conditional 99\%-confidence interval.}
 \end{figure}

In the case of the windowing method, the PSD estimates are affected by leakage, and follow the distorted periodogram. This is visible for both gap patterns A and B. The lowest frequencies are more affected by spectral leakage, and the PSD estimates are accurate in a larger frequency band for five-day periodic gaps than for Daily random gaps. If we consider the DA method, the PSD estimates are consistent with the true PSD within error bars. This implies that the gap data imputation step allows to recover the right statistic of the noise and removes the bias caused by leakage. We can also check the consistency of the imputation in the time domain, by looking at one missing data draw such as in Fig.~\ref{fig:time_imputation}.

In addition, the right-hand side plot in Fig.~\ref{fig:psd_estimation} corroborate the results presented in Sec.~\ref{sec:parameter_results} for $f_0~=~0.1$ mHz and Daily random gaps where we obtain a quasi-flat posterior distribution. The leakage effect is so large that the signal frequency peak (light orange) is blurred into the noise, making the MCMC algorithm fail to recover the parameters. However, if we consider the 2 other peaks in orange shades corresponding to higher signal frequencies, the larger the frequency, the more they come out of the noise.

\subsection{\label{sec:parameter_results}Results of parameter estimation for 2 sources}

Given that LISA will observe tens of thousands of sources at the same time, an important aspect of data analysis is the ability to distinguish two sources whose frequency bandwidths overlap. Therefore we want to assess our ability to resolve two galactic binaries with close frequencies, \textit{i.e} our ability to determine the right number of sources in a given bandwidth. To this aim, we simulate a data set with a first source whose frequency is $f_1 = 0.2$ mHz and a second source whose frequency is slightly offset with respect to the first one: $f_2 = f_1 + \Delta f$. We study the cases where $\Delta f = 10^{-n}$ Hz, with $n = 7,\, 8,\, 9,\, 10$. We assume different sky positions and distances, as indicated in Table~\ref{tab:2sources}.
 
 \begin{table}
 \begin{ruledtabular}
 \begin{tabular}{l c c}
Parameter & Source 1 & Source 2 \\
\hline
Amplitude [strain $\times 10^{-20}$] & 2 & 2\\ 
Frequency [mHz] & $0.2$ & $0.2 + \Delta f$ \\
Ecliptic latitude [rad] & 0.47 & -1.0\\
Ecliptic longitude [rad] & 4.19 & 5.5\\
Inclination [rad] & 0.179 & 0.1\\
Initial phase [rad] & 5.78 & 2.89\\
Polarization [rad] & 3.97 & 0.21\\
 \end{tabular}
 \end{ruledtabular}
  \caption{\label{tab:2sources}Characteristics of the 2 simulated sources in each data stream. Four data streams are generated where the sources are 100, 10, 1 or 0.1 nHz apart in frequency, with different sky location and orientation.}
 \end{table} 
 
Even if dedicated detection and estimation algorithms allowing to determine the dimension of the parameter space have been developed~\cite{Littenberg2011}, here we adopt a simple approach where we perform two estimations: one assuming a single source in the signal model, the other assuming two sources. In the latter assumption we set the constraint $f_1 < f_2$ in the frequency prior, which allows the MCMC algorithms to better cluster the posteriors (avoiding frequent jumps between two modes). The 1-source and 2-source model estimations are done in the case of complete data and in the case of gapped data, using the windowing method and then the DA method. In order to assess the validity of the model (including the priors on its parameters, see \citet{Sharma2017}), at the end of each estimation we estimate the Bayes factor
\begin{eqnarray}
B_{21} \equiv \frac{p\left( \boldsymbol{y} | \mathcal{M}_{2} \right) }{p\left( \boldsymbol{y} | \mathcal{M}_{1} \right) },
\end{eqnarray}
where $p\left( \boldsymbol{y} | \mathcal{M}_{2} \right)$ is the evidence associated with the posterior distribution for model $\mathcal{M}_{i}$ with $i$ source(s), and is  defined as
\begin{eqnarray}
p\left( \boldsymbol{y} | \mathcal{M}_{i} \right) = \int_{\Theta} p\left( \boldsymbol{\theta},  \mathcal{M}_{i} \right) p\left( \boldsymbol{y} |  \boldsymbol{\theta}, \mathcal{M}_{i} \right).
\end{eqnarray} 
The evidences are computed via thermodynamic integration~\cite{Ogata1989, Gelman1998, Lartillot2006}, a method which uses the results of the PTMCMC algorithm at all temperatures, and which has already been applied in GW data analysis~\cite{Littenberg,Robson2018}. Details about our implementation of this method are given in Appendix~\ref{sec:thermo}.

In Fig.~\ref{fig:2sources_freq} we gather the estimated posterior distributions of source frequencies $\hat{f}_{1}$ and $\hat{f}_{2}$ for each injected frequency separation. In the complete data case (solid black curve), the frequencies are well resolved for $\Delta f > 1$ nHz, where the two posterior distributions start to be superimposed, as their standard deviations is about 2.5 nHz. 

 \begin{figure*}[ht]
 \centering
 \begin{tabular}{c c c c c}
 \rotatebox{90}{\hspace{0.5cm}Five-day periodic gaps} 
 &
 \includegraphics[width=0.24\textwidth]{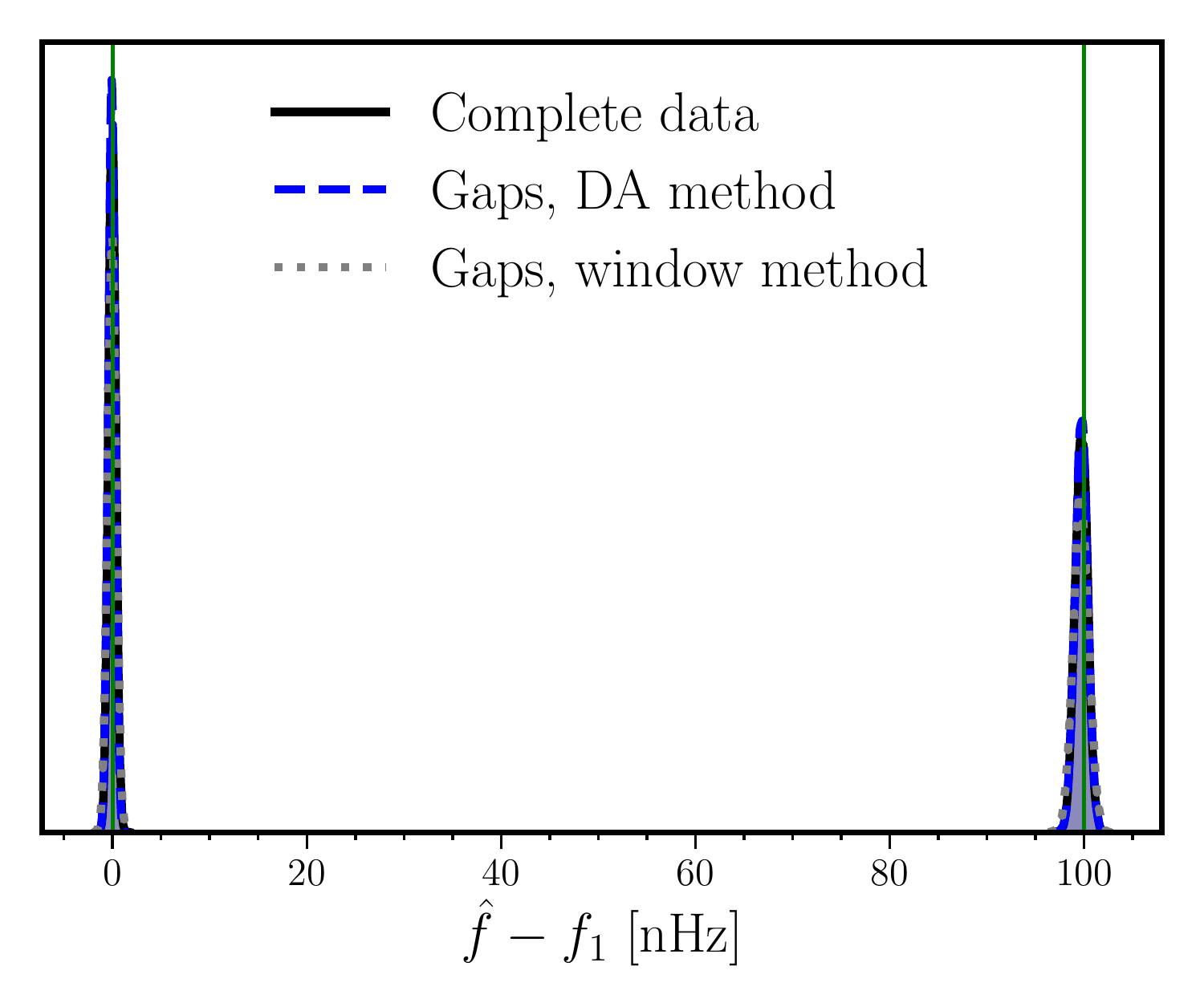}%
& 
 \includegraphics[width=0.24\textwidth]{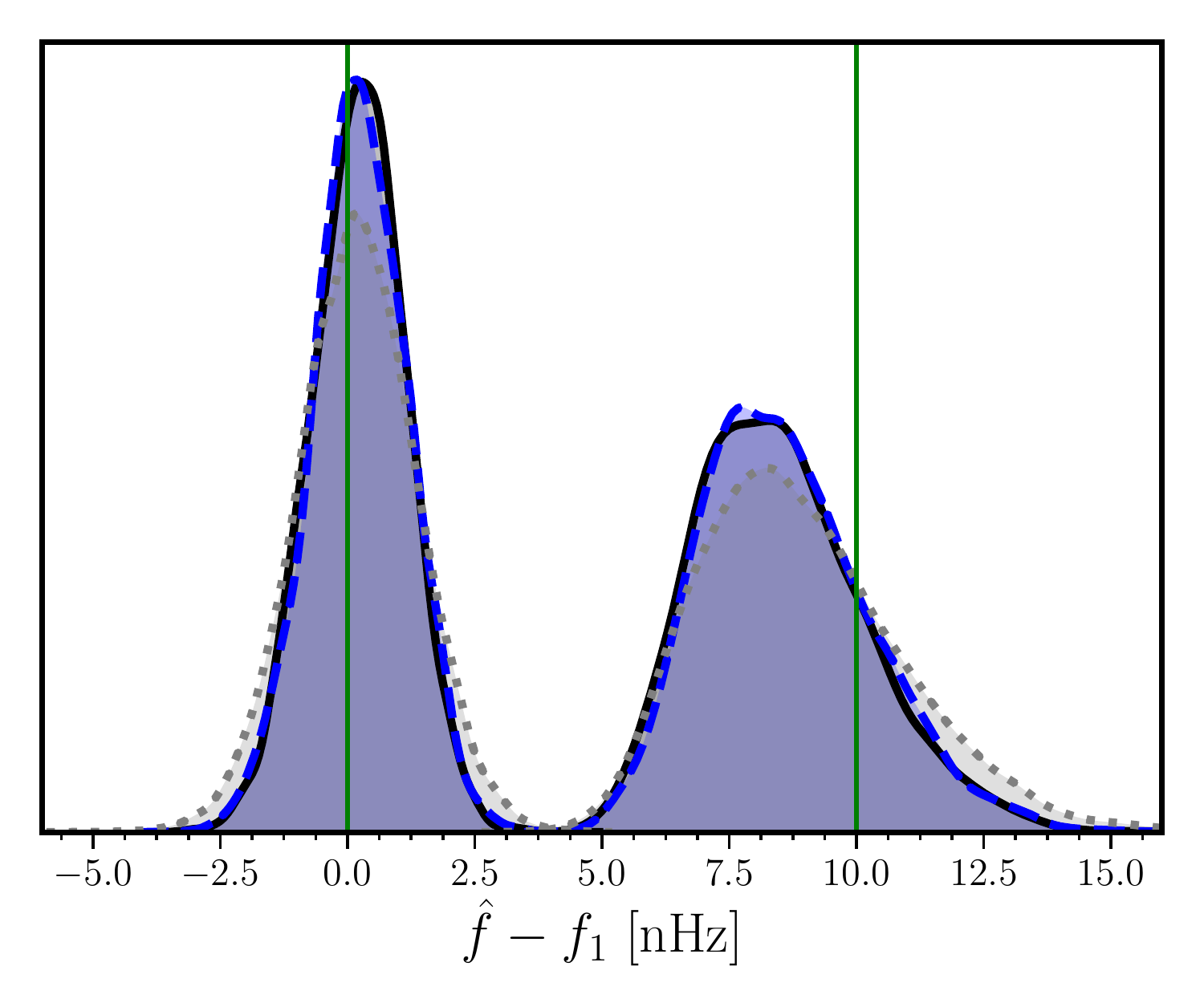}
 &
  \includegraphics[width=0.24\textwidth]{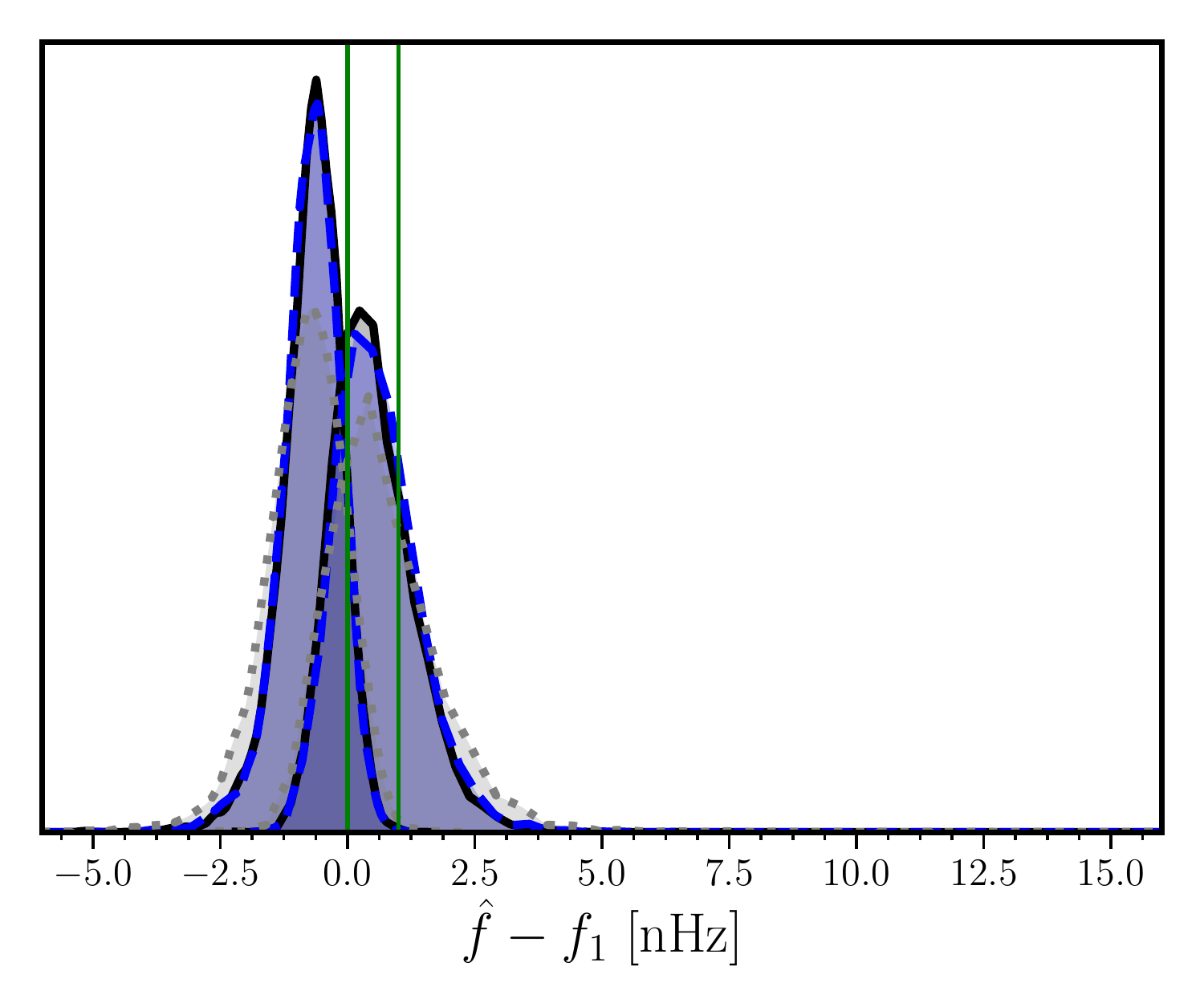} 
   &
  \includegraphics[width=0.24\textwidth]{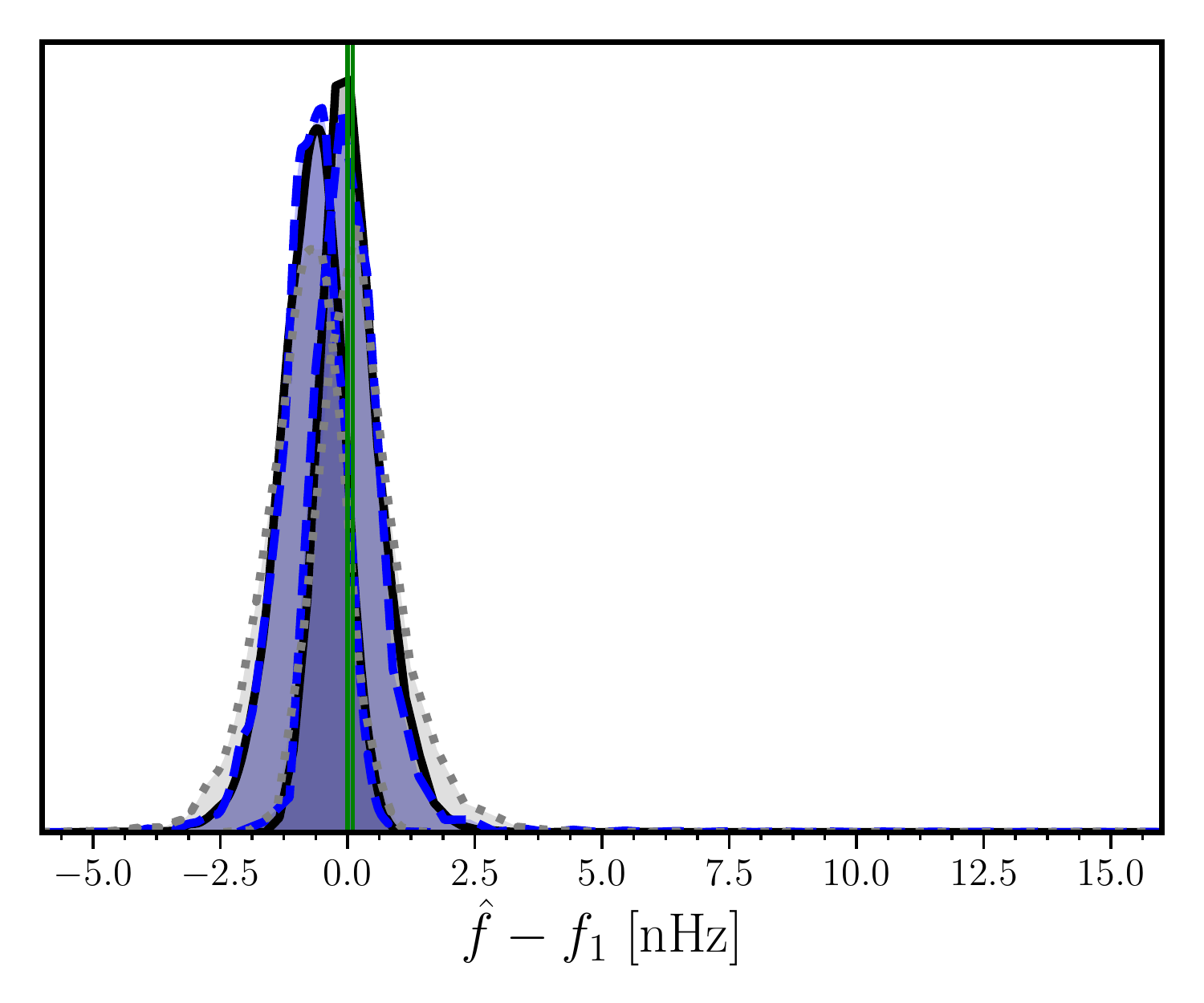} \\ 
  \rotatebox{90}{\hspace{0.7cm}Daily random gaps} 
  &
 \includegraphics[width=0.24\textwidth]{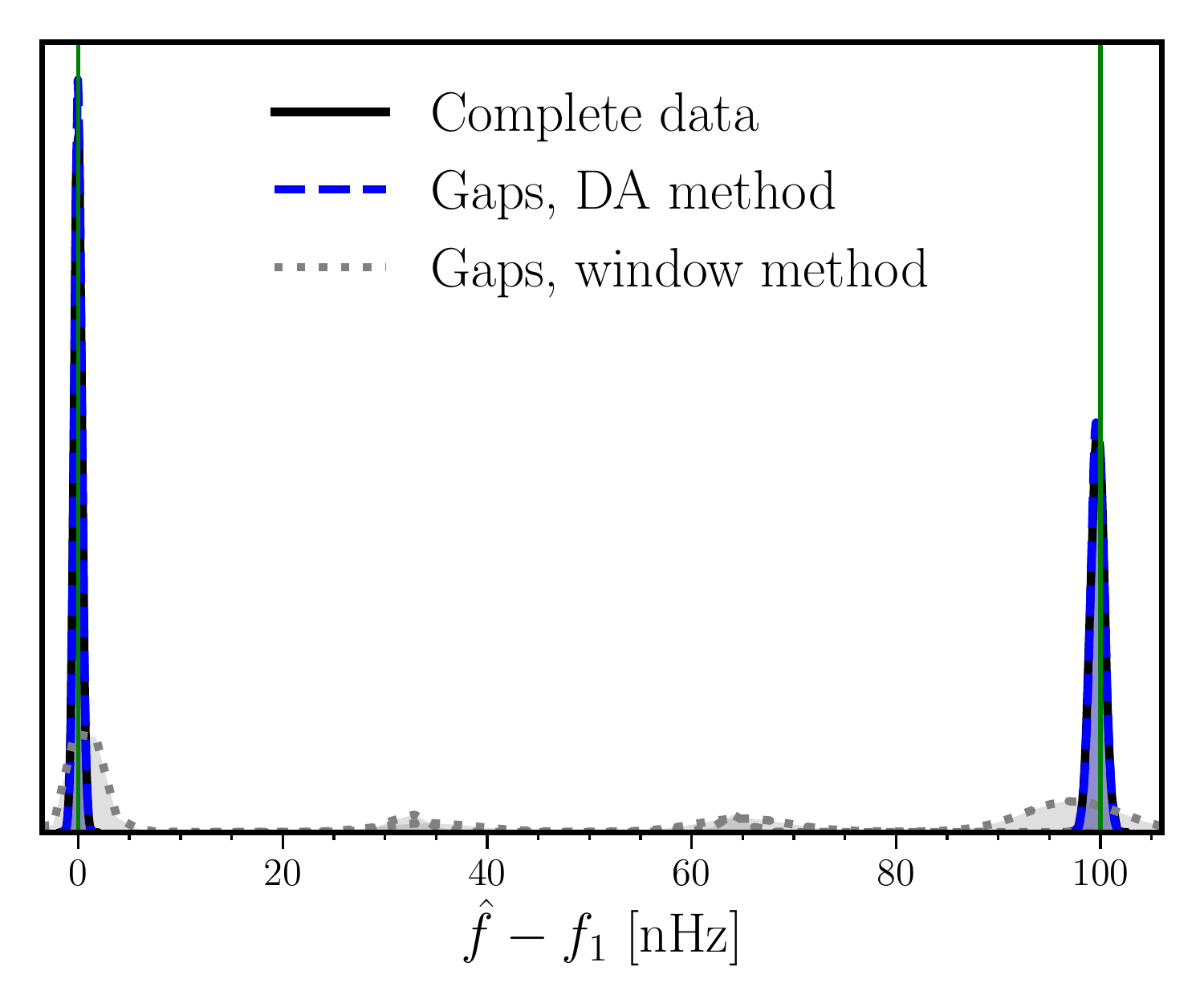}%
& 
 \includegraphics[width=0.24\textwidth]{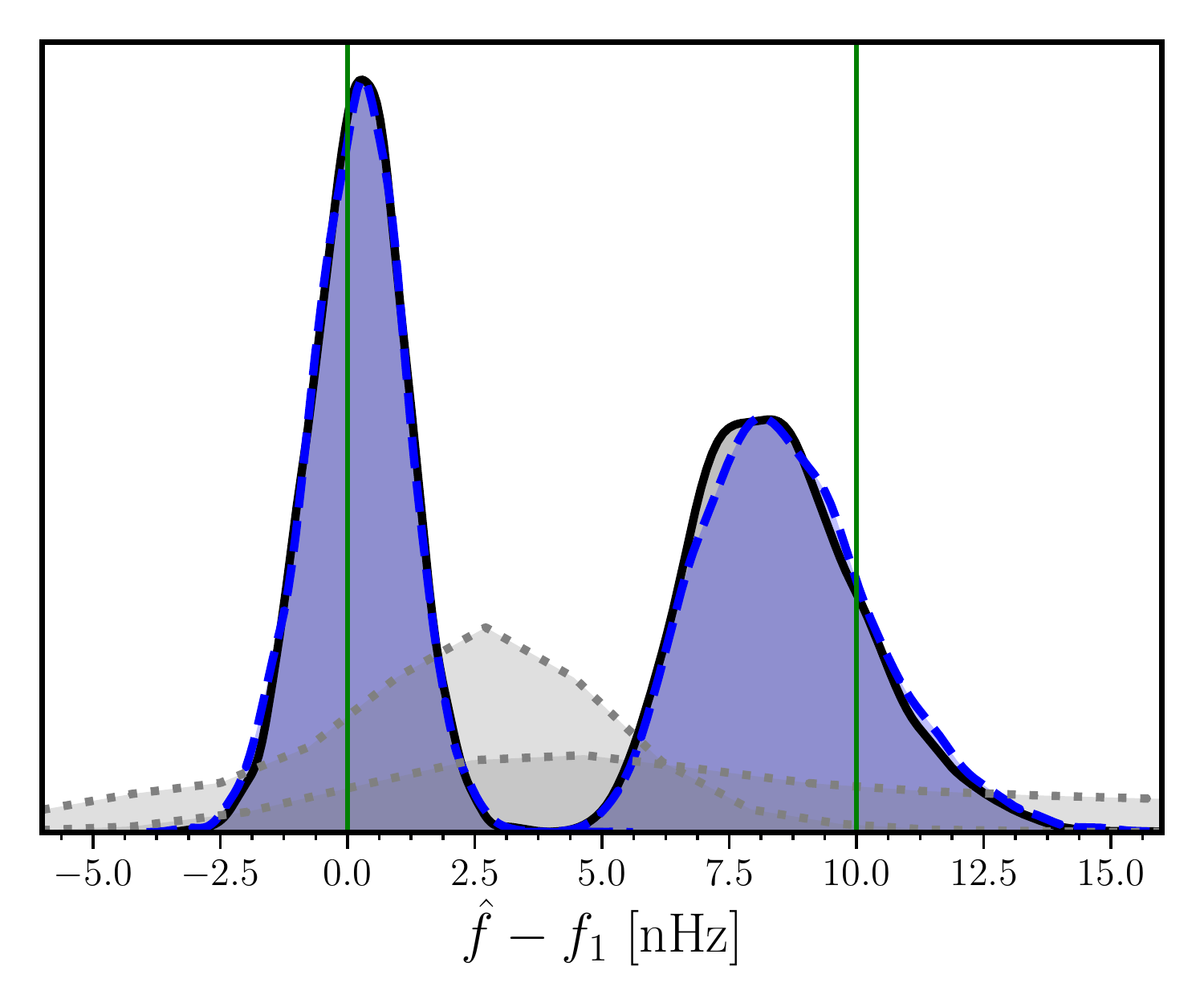}
 &
  \includegraphics[width=0.24\textwidth]{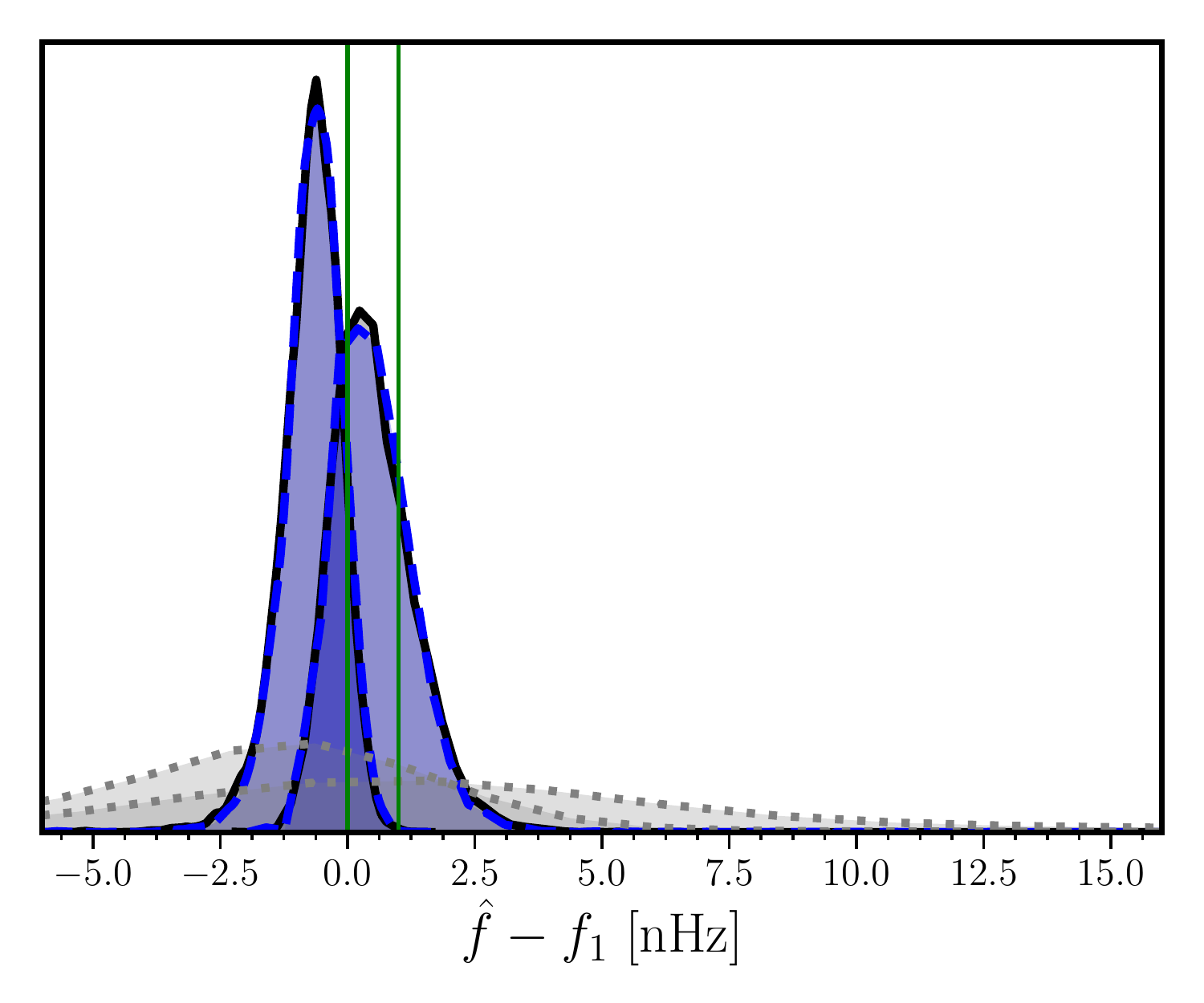} 
   &
  \includegraphics[width=0.24\textwidth]{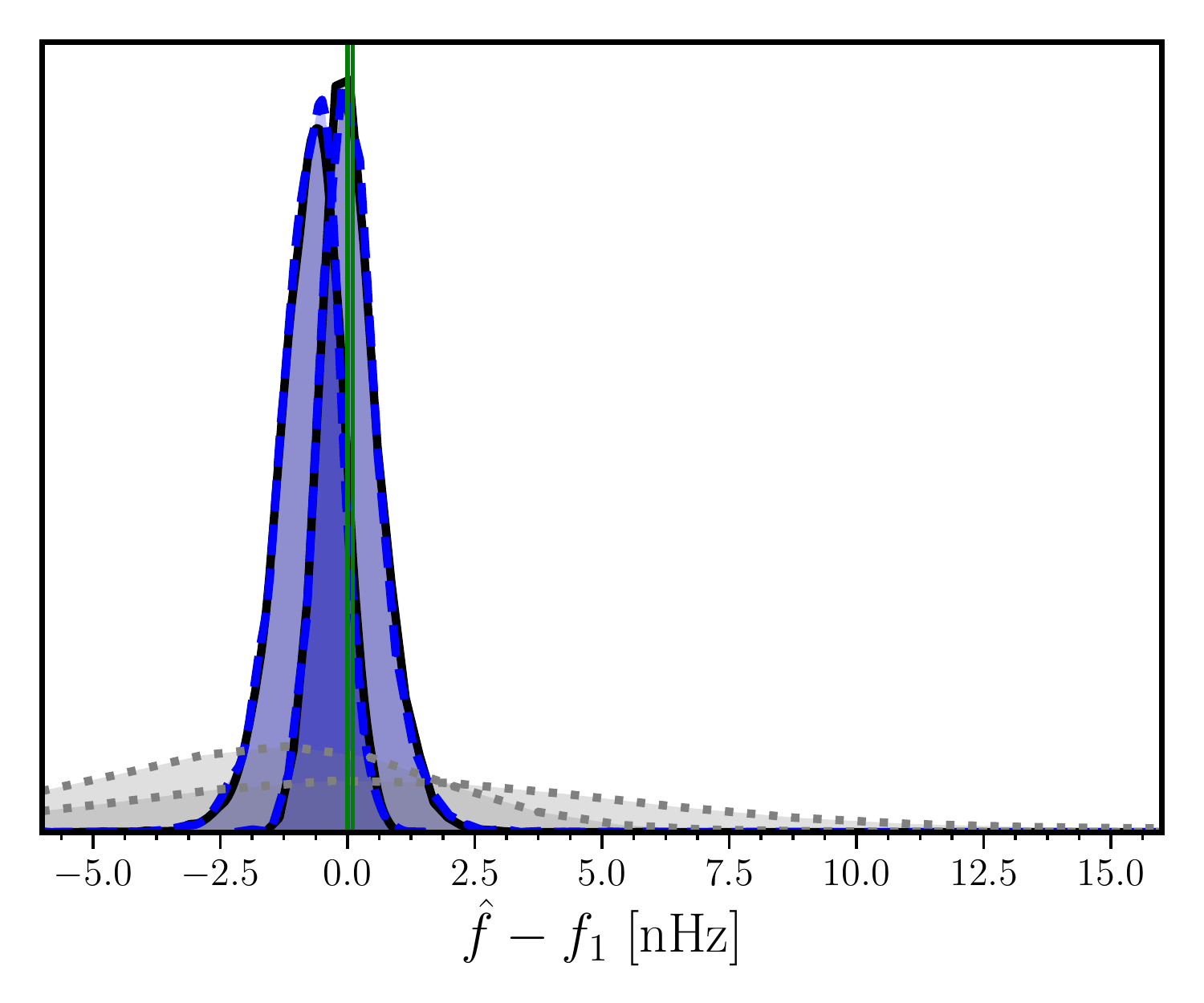} \\
 & $\Delta f = 10^{-7}$ Hz &  $\Delta f = 10^{-8}$ Hz & $\Delta f = 10^{-9}$ Hz & $\Delta f = 10^{-10}$ Hz
 \end{tabular}
 \caption{\label{fig:2sources_freq}Posterior distribution of frequencies $\hat{f}_1$ and $\hat{f}_2$ of two sources with different sky locations, obtained with the 2-source model. We show 4 different cases of frequency separation $\Delta f$ (which is decreasing from left to right) with and without random gaps. The sample values are offset by $f_1 = 2$ mHz for clarity. Kernel estimates of the posterior distribution densities in the case of complete data are represented by black solid curves; the case of gapped data with the use of the windowing method is represented by the dotted gray curves, and the case of gapped data with the use of the DA method is represented in dashed blue. Vertical green lines indicate the true value of both frequencies.} 
 \end{figure*}

In the case of periodic gaps, this behavior is observed for both windowing and DA methods, although the statistical error increases by about 30 \% in the case of the windowing method. In the case of random gaps, the posteriors are much more spread when using the windowing method, making impossible to resolve the frequencies for separations of 10~nHz and below. The posteriors obtained with the DA method are very similar to the complete data case, restoring the frequency resolution power to a level comparable with full-data resolution.

\begin{figure*}[ht]
 \centering
 \begin{tabular}{cc} 
  \hspace{0.8cm}  Five-day periodic gaps & \hspace{0.8cm} Daily random gaps \\
  \includegraphics[width=0.45\textwidth]{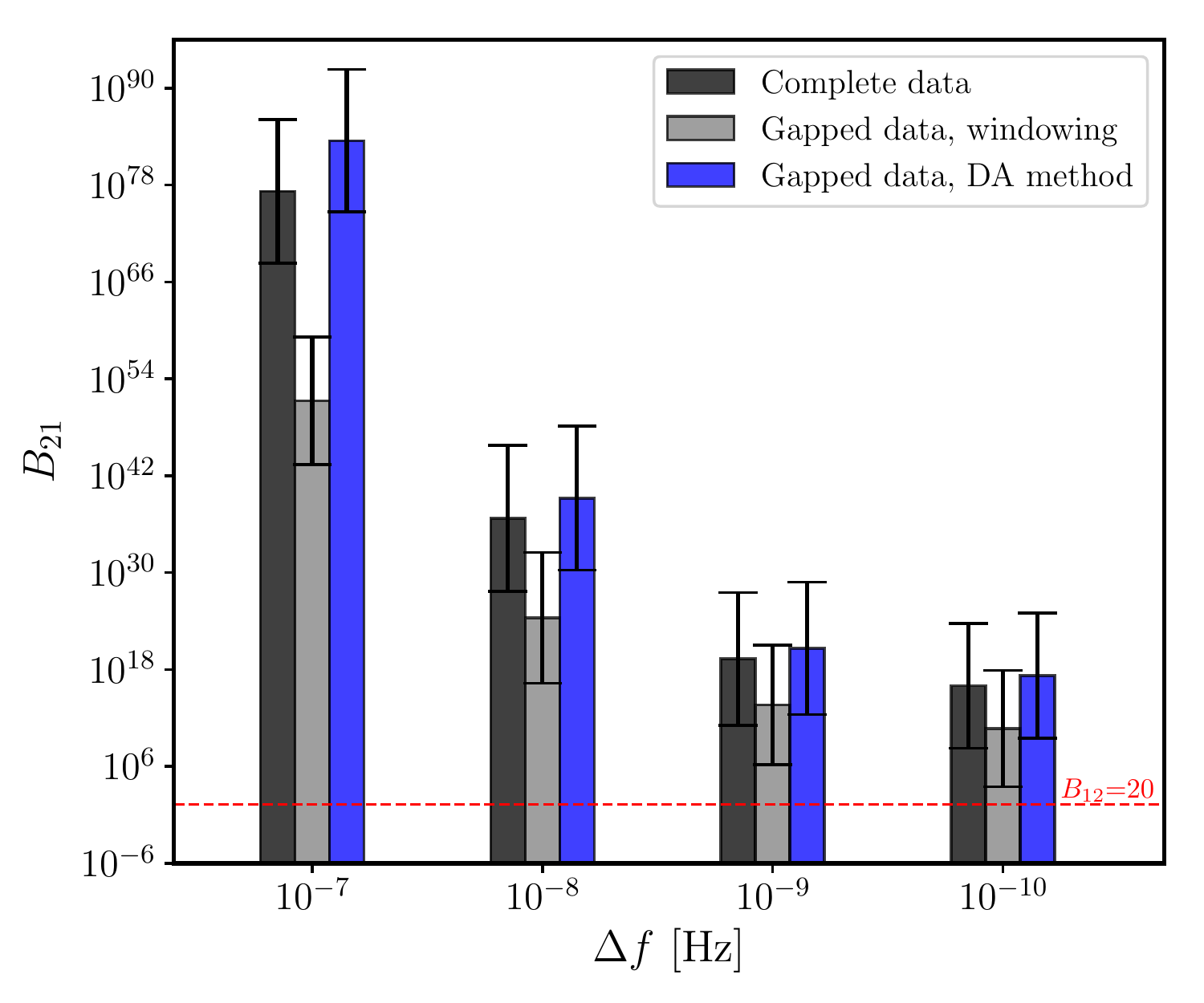} 
  &
   \includegraphics[width=0.45\textwidth]{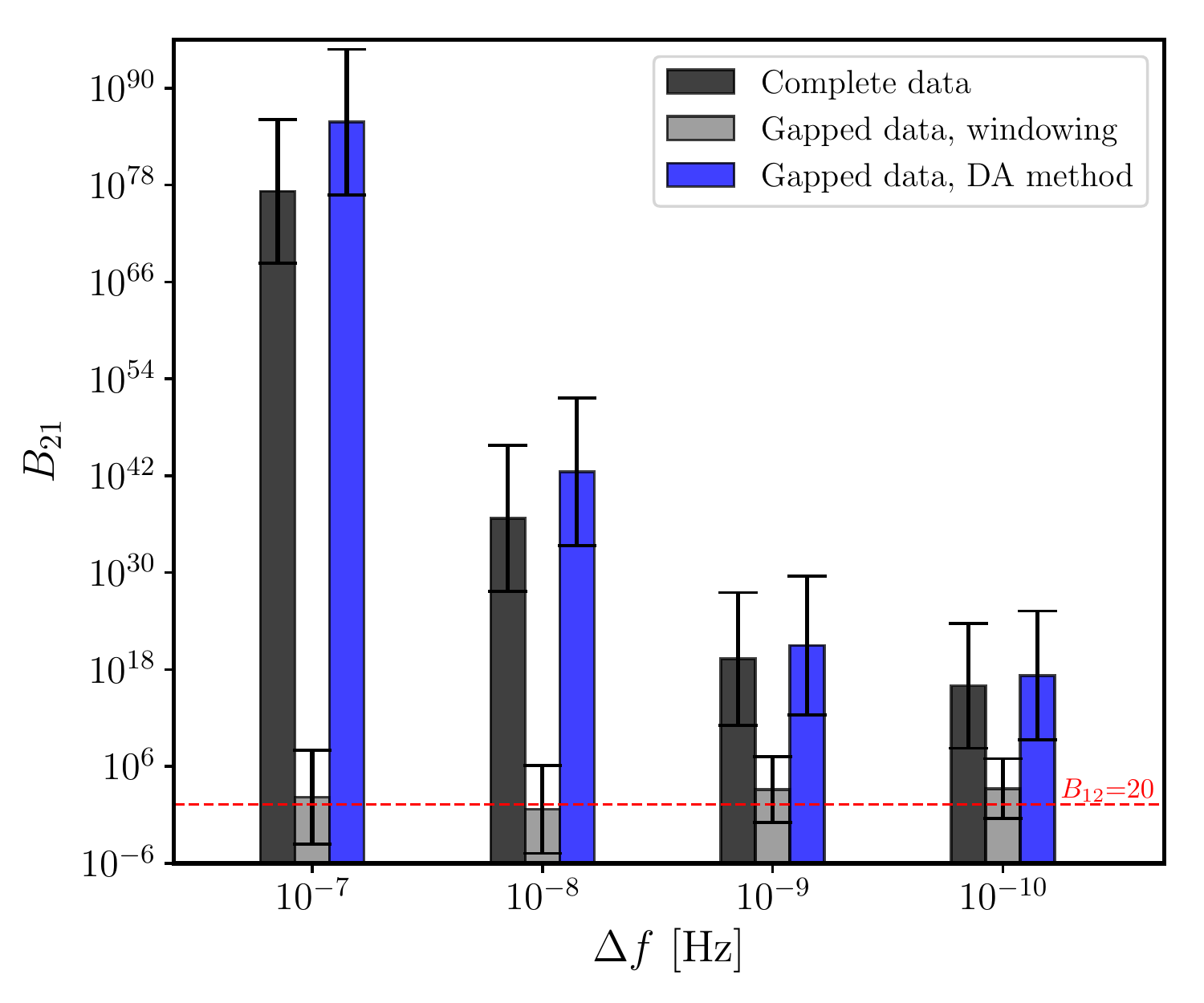} 
 \end{tabular}
 \caption{\label{fig:bayes_factors}Estimated Bayes factors $B_{21}$ as a function of the frequency separation $\Delta f = f_2 - f_1$ between the two sources with different sky locations, in the case of five-day periodic gaps (left-hand side) and random daily gaps (right-hand side). In both panels the posterior obtained from complete data is shown in solid black for reference, while the posteriors obtained from gapped data with the windowing method are shown in dotted gray, and in dashed blue with the DA method. The dashed red horizontal lines corresponds to the threshold above which the Bayes factor indicates positive evidence towards 2 sources (values suggested in \citet{Romano2017}). The error bar account for the possible bias due to the numerical integration.}
 \end{figure*}

The frequency estimates can be compared to the values obtained for the estimated Bayes factors, plotted in Fig.~\ref{fig:bayes_factors}. As in Fig.~\ref{fig:2sources_freq}, they are ordered by decreasing separations between the 2 source frequencies injected in the simulated data. We show the case of a complete data series (black vertical bars), along with gapped data with the windowing method (gray) and the DA method (blue). The top and bottom panels correspond to periodic and random gaps respectively. For periodic gaps, although the windowing method yields smaller values of $B_{21}$ than the DA method, the Bayes factor significantly favor a 2-source model, both in the case of complete and masked data, regardless of the method used. For $\Delta f = 0.1$ nHz the value that we compute with the windowing method gets closer to the positive threshold (indicated by the red dashed horizontal line) which we set to $B_{21} = 20$ \cite{Romano2017}. This suggests that for such a frequency separation it would be difficult to discriminate the two models if the SNR was smaller (as the amplitudes of Bayes factors would be lower). Besides, not surprisingly, the values of the Bayes factors are decreasing as the frequencies of the sources get closer.

If we now look at the case of daily random gaps in the bottom panel, we see that in most cases the windowing method yields values of $B_{21}$ lying below or close to the positive threshold, and does not allows us to conclude on a preference model given the large systematic error bars. However, the DA method allows us to confidently favor the presence of 2 sources, as does a complete data series. They both give $B_{21} > 150$ which corresponds to a very strong positive test. This means that even when the frequencies cannot be distinguished, we are still able to tell whether there is one or two sources in the data. Besides, one may remark that the Bayes factors computed with the DA method seem to have a systematically larger amplitude than for the case of full data, although the error bars suggest that this could only be coincidental for this particular realization of the data. The difference can be explained by the fact that the two posterior distributions that are probed in the case of complete and missing data are not the same, since in the former case we sample for $p\left( \boldsymbol{\theta} | \boldsymbol{y} \right)$ and in the latter we sample for $p\left( \boldsymbol{\theta} , \boldsymbol{y}_{m} |  \boldsymbol{y}_{o} \right)$ which includes the missing data points as additional parameters in the model. Although increasing the sampling cadence of missing data tends to decrease this systematic, we may have to define a more adapted threshold when using the DA method, which will be investigated in future studies.

\section{\label{sec:conclusions}Conclusions}

We tackled the problem of GW parameter estimation in the presence of data gaps in LISA measurements, by (i) assessing their impact on the performance of a standard method using window functions in the time domain, and (ii) by minimizing this impact through the development of an adapted Bayesian method. 

For the assessment task, we introduced a figure of merit that we call effective signal-to-noise ratio, which takes into account the signal and noise leakage effect that is inherent in any finite-time windowing method. Our study focused on compact galactic binary sources with different frequencies, and on 2 kinds of gap pattern: weekly, pseudo-periodic gaps were generated to mimic possible satellite antenna repointing operations, and daily random gaps were generated to account for a situation where frequent and loud instrumental glitches would corrupt some data, making them unusable. We showed that short and frequent data gaps (glitch-masking type) are more impacting than longer and fewer gaps (antenna-type). They can degrade the effective SNR by up to 70\% for low frequency (0.1 mHz) sources, with only 1\% data losses. This is explained by the large stochastic noise power leakage that is transferred to low frequency when masking (even smoothly) the data. We also showed that this degradation gets larger when the frequency of the source decreases. 
The precision of GW parameter estimation is found to be in direct correlation with the noise leakage, and drops by the same amount as the effective SNR. We also studied how the resolution power depends on gaps. 
For weekly gaps, we showed that time-windowing maintain a good performance for frequency resolution and source disentanglement, with a slight increase in the statistical error, on the order of a few percents. However, daily gaps largely deteriorate the ability to distinguish between two sources that are close in frequency.

In order to mitigate the loss of effective SNR due to time windowing, we developed an alternate Bayesian method to handle data gaps in a statistically consistent framework, called data augmentation. We introduce an extra step in the sampling of the posterior distribution where, beside GW and noise PSD parameters, we sample for missing data. 
This allows us to use standard MCMC sampling techniques on iteratively reconstructed time series, which removes leakage effects while properly taking data gaps into account in the posterior distribution of GW parameters. 
We showed that in cases where the time-windowing treatment of gaps performs poorly, data augmentation restores UCB parameter precision at a level that is comparable with complete data analysis, thereby increasing the ability to distinguish between different sources. One advantage of this method is that it is agnostic to any waveform model, and can be easily plugged in any existing MCMC scheme. By limiting the sampling cadence of missing data, this "plug-in" represents an moderate extra computational cost, lower than 5\%. 
Furthermore, for a given fraction of data losses, the method shows equivalent performance with different gap patterns. 

As a conclusion, our study shows that data gaps may degrade LISA's performance if not properly handled, impacting both parameter precision and source resolution power. Handling them properly presents a computational challenge, but we proposed a Bayesian data augmentation inference method that efficiently tackles the problem in a statistically consistent way, and gives promising results. 
While the quasi-monochromatic sources chosen for our study allows us to easily assess the dependence of gap impact on frequency, the next obvious step that will be taken by future works is to apply the DA method to other LISA sources such as massive black hole binaries, which span larger frequency ranges. In addition, some limitations of the method remain to be addressed. In particular, the computational complexity of the missing data imputation step is currently larger than GW posterior sampling steps. While we showed that reducing the imputation cadence allows us to circumvent this problem while maintaining good sampling accuracy, better sampling may be needed for more complex problems. We plan to develop a more efficient algorithm in the future, by using sparse approximation of matrices and parallel computations.

\appendix

\section{\label{sec:modulation_functions}Expression of modulation functions}

Here we give the expressions of the functions $F_{+}$, $F_{\times}$ that are involved in Eq.~(\ref{eq:phase_shift_arm}) describing LISA's response to GWs. They are given by
\begin{eqnarray}
\label{eq:modulation_factors_def}
F^{i}_{+}\left(  t\right) &\equiv &  \cos(2\psi)  u^{i}(t) -  \sin(2\psi) v^{i}(t) ;\nonumber \\
F^{i}_{\times}\left( t\right) & \equiv & \sin(2\psi) u^{i}(t) + \cos(2\psi) v^{i}(t),
\end{eqnarray}
where $u^{(i)}$ and $v^{(i)}$ are modulation functions  If $\boldsymbol{n}_{\theta}$, $\boldsymbol{n}_{\phi}$ and $\boldsymbol{k}$  are the orthonormal vectors of the  observational coordinate system in the SSB frame, these modulation functions are equal to 
\begin{eqnarray}
u^{i}(t) & = & \frac{\pi \nu_0 L_i}{c} \left[ \left( \boldsymbol{n}_{\theta} \cdot \boldsymbol{n}_i \right)^2 - \left( \boldsymbol{n}_{\phi} \cdot \boldsymbol{n}_i \right)^2 \right];   \nonumber \\
v^{i}(t) & = & \frac{\pi \nu_0 L_i}{c} \left( \boldsymbol{n}_{\theta} \cdot \boldsymbol{n}_i \right) \left( \boldsymbol{n}_{\phi} \cdot \boldsymbol{n}_i \right).
\end{eqnarray}

\section{\label{sec:fourier_series}Fourier-domain model}

In this section we derive an approximate Fourier-domain model for the UCB. To do that, it is convenient to expand the waveform response in Fourier series.
We start from Eq.~(\ref{eq:phase_linear_time_model}) that we reproduce here for convenience:
\begin{eqnarray}
\label{eq:phase_linear_time_model2}
\Delta \Phi^{(i)} (t) = \sum_{j=1}^{4} a^{(i)}_{j} g_{j}(t).
\end{eqnarray}
The elementary functions $g_{j}$ of this combination can be written as:
\begin{eqnarray}
\label{eq:g_functions}
g_{1}(t) & = &  u^{(i)}(t) \cos \Phi_s\left( t - d(t) \right) \nonumber \\
g_{2}(t) & = &  v^{(i)}(t) \cos \Phi_s\left( t - d(t) \right) \nonumber \\
g_{3}(t) & = &  u^{(i)}(t) \sin \Phi_s\left( t - d(t) \right)  \nonumber \\
g_{4}(t) & = &  v^{(i)}(t) \sin \Phi_s\left( t - d(t) \right).
\end{eqnarray}
At this point we can develop both the modulation functions $u^{(i)}(t)$ and $u^{(i)}(t)$ and the delayed GW phase $\Phi_s\left( t - d(t) \right)$ in Fourier series. The former can be expressed as a sum of sinusoidal functions with integer multiples of the orbital angular frequency $\omega_T = 2 \pi /T$:
\begin{eqnarray}
\label{eq:modulation_harmonic_decomp}
u^{i}(t) & = & \sum_{m=0}^{4} U^{(i)}_{c,m} \cos\left( m \omega_T t \right) + U^{(i)}_{s,m} \sin\left( m  \omega_T t  \right) \nonumber \\
v^{i}(t) & = & \sum_{m=0}^{4} V^{(i)}_{c,m} \cos\left( m \omega_T t \right)  + V^{(i)}_{s,m} \sin\left( m  \omega_T t  \right).
\end{eqnarray}  
Then the modulated cosine and sine functions of the GW phase can be expanded as
\begin{eqnarray} 
\label{eq:jacobi-anger}
e^{j\Phi_s\left(t-d(t)\right)} & \approx & e^{j\Phi_s\left(t \right) + j \omega_0 \tau_0 \cos\left(\phi - \omega_T t \right) } \nonumber \\
& =& \sum_{n=-\infty}^{+\infty} J_{n}\left( \omega_0 \tau_0 \right) j^{n}   e^{j\Phi_s\left(t \right) + jn\left(\phi - \omega_T t \right) }, 
\end{eqnarray}
where in the first line we applied the slow chirp approximation and set $\omega_0 \equiv 2 \pi f_0 \; ; \; \tau_0  \equiv \frac{R \sin \theta}{c} $, and in the second line we used the Jacobi - Anger identity. 

The Fourier transform of the above expression is a convolution between the exponential harmonic at angular frequency $\omega_T$ and the Fourier transform of the GW exponential phase, that we write as
\begin{eqnarray}
\label{eq:exponential_phase_ft}
\tilde{v}_T(f)  = \int_{-\infty}^{+\infty} \Pi_{0,T}(t) e^{-j\Phi_s(t)} e^{-2 j\pi f t} dt,
\end{eqnarray}
where $\Pi_{0,T}(t)$ is the rectangular window of size $T$, which accounts for the fact that we observe the signal during a finite duration. In the case of a slowly chirping wave with frequency derivative $\dot{f}_{0}$, Eq.~(\ref{eq:exponential_phase_ft}) can be expressed as 
\begin{eqnarray}
\label{eq:chirping_v}
 \tilde{v}_T(f) & = & I_T(f + f_0) - I_0(f + f_0); \text{ with:} \nonumber \\
  I_{t}(f) & \equiv & \frac{e^{j \pi \left( \frac{1}{4} + \frac{f^2}{\dot{f}_0} \right) } }{2\sqrt{\dot{f}_0}} \mathrm{erfi}\left( \sqrt{\frac{\pi}{\dot{f}_0}} e^{j\frac{\pi}{4}} \left( \dot{f}_0 t + f\right) \right) 
\end{eqnarray}
where $\mathrm{erfi}$ is the imaginary error function.
When chirping can be neglected, Eq.~(\ref{eq:exponential_phase_ft}) reduces to a cardinal sinus function
\begin{eqnarray}
\label{eq:mono_v}
\tilde{v}_T(f) = T e^{-j \pi (f + f_0) T} \sinc \left[ (f + f_0) T) \right].
\end{eqnarray}

Let us now take the Fourier transform of Eqs.~(\ref{eq:jacobi-anger}) after truncating the series to some order $n_c$:
\begin{eqnarray} 
\label{eq:jacobi-anger_freq}
\mathcal{F}\left[e^{j\Phi_s\left(t-d(t)\right)} \right](f) \approx \sum_{n=-n_c}^{+n_c} J_{n}\left( \omega_0 \tau_0 \right) j^{n} e^{ j n \phi} \tilde{v}_T(f - n f_T). \nonumber
\end{eqnarray}
Let us define $y_c(t) \equiv \cos \Phi_s\left( t - d(t) \right)$ and $y_s(t) \equiv \sin \Phi_s\left( t - d(t) \right)$, which can also be written as
\begin{eqnarray}
y_c(t) &=& \frac{1}{2} \left[ e^{j\Phi_s\left(t-d(t)\right)} + e^{-j\Phi_s\left(t-d(t)\right)}\right], \nonumber \\
y_s(t) &=& \frac{1}{2j} \left[ e^{j\Phi_s\left(t-d(t)\right)} - e^{-j\Phi_s\left(t-d(t)\right)}\right].
\end{eqnarray}
Hence the Fourier transform of $y_c(t)$ and $y_s(t)$ are
\begin{eqnarray}
\label{eq:y_c_y_s_freq}
\tilde{y}_c(f) &=& \frac{1}{2}  \sum_{n=-n_c}^{+n_c}   J_{n}\left( \omega_0 \tau_0 \right)   \Big[ e^{ j n \left(\phi + \frac{\pi}{2} \right)} \tilde{v}_T(f - n f_T) \nonumber \\
&& +e^{- j n \left(\phi + \frac{\pi}{2} \right)} \tilde{v}^{*}_T(- f + n f_T)  \Big], \nonumber \\
\tilde{y}_s(f) &=& \frac{1}{2j}  \sum_{n=-n_c}^{+n_c}   J_{n}\left( \omega_0 \tau_0 \right)   \Big[ e^{ j n \left(\phi + \frac{\pi}{2} \right)} \tilde{v}_T(f - n f_T) \nonumber \\
&& - e^{- j n \left(\phi + \frac{\pi}{2} \right)} \tilde{v}^{*}_T(- f + n f_T)  \Big].
\end{eqnarray}
To obtain a more compact expression, we can also restrict the frequency domain to positive frequencies, since the likelihood are usually computed for positive frequencies only. According to Eq.~(\ref{eq:mono_v}) the function $\tilde{v}_T(f)$ is centered in $-f_{0}$, so that only terms of the form $\tilde{v}_T\left(-f + k f_T\right)$ are dominant for $f > 0$:
\begin{eqnarray}
\tilde{y}_c(f) &\approx & \frac{1}{2}  \sum_{n=-n_c}^{+n_c}   J_{n}\left( \omega_0 \tau_0 \right)  e^{- j n \varphi} \tilde{v}^{*}_T(- f + n f_T), \nonumber \\
\tilde{y}_s(f) & \approx & j \tilde{y}_c(f),
\end{eqnarray}
where we set $\varphi \equiv \phi + \frac{\pi}{2}$. 

With this result in hands, the Fourier transform of the functions $g_j$ in Eq.~(\ref{eq:g_functions}) can be computed from the convolution of Eq.~(\ref{eq:jacobi-anger}) with the Fourier transform of Eq.~(\ref{eq:modulation_harmonic_decomp}). For example, for $j=1$ we have $g_{1}(t) = u^{(i)}(t) y_c(t)$, hence for $f>0$ we get
\begin{eqnarray}
\tilde{g}_{1}(f) & = & \int_{-\infty}^{+\infty} \tilde{u}^{(i)}\left( f - f' \right) \tilde{y}_c\left( f'\right) df' \nonumber \\
& = & \frac{1}{4}\sum_{n=-n_c}^{+n_c} \sum_{m=0}^{4}  J_{n}(\omega_0 \tau_0) e^{- j n \varphi} \left( U^{(i)}_{c,m}  - j U^{(i)}_{s,m}  \right)\nonumber \\
&&  \Big[ \tilde{v}^{*}_T\left(- f + (n + m) f_T \right) + \tilde{v}^{*}_T\left(- f + (n - m) f_T \right) \Big]. \nonumber
\end{eqnarray}
Similar expressions are obtained for $\tilde{g}_j(f),\, j=2, 3, 4$ which gives us an explicit formula for the frequency-domain waveform response (\ref{eq:phase_linear_time_model2}).

\section{\label{sec:thermo}Implementation of thermodynamic integration}

In this section we details the implementation that we use to calculate the source model evidence. If we consider a particular model $\mathcal{M}$, we can define a tempered version of the evidence:
\begin{eqnarray}
Z\left( \beta \right) \equiv \int_{\Theta}  q_{\beta}\left( \boldsymbol{\theta} \right) d\boldsymbol{\theta},
\end{eqnarray}
where 
\begin{eqnarray}
q_{\beta}\left( \boldsymbol{\theta} \right) = p\left( \boldsymbol{y} | \boldsymbol{\theta} , \mathcal{M} \right)^{\beta} p\left( \boldsymbol{\theta} | \mathcal{M} \right),
\end{eqnarray}
and $\beta$ is the inverse temperature. Thus we want to estimate $Z\left( 1 \right)$. One can show that \cite{Lartillot2006} 
\begin{eqnarray}
\label{eq:log_like_exp}
\frac{\partial \log Z(\beta)}{\partial \beta} & = & \mathrm{E}_{\beta}\left[ \frac{\partial \log q_{\beta}(\boldsymbol{\theta}) }{\partial \beta} \right] \nonumber \\
&=& \mathrm{E}_{\beta}\left[ \log p\left( \boldsymbol{y} | \boldsymbol{\theta} , \mathcal{M} \right) \right],
\end{eqnarray}
where the expectation is taken, for fixed $\beta$, with respect to the tempered posterior distribution.
We can derive the evidence by taking the integral of Eq.~(\ref{eq:log_like_exp}) along all temperatures:
\begin{eqnarray}
\label{eq:evidence_integral}
\log Z(1) - \log Z(0) = \int_{0}^{1}\mathrm{E}_{\beta}s\left[ \log p\left( \boldsymbol{y} | \boldsymbol{\theta} , \mathcal{M} \right) \right] d\beta,
\end{eqnarray}
where $Z(0) = \int_{\Theta}  p\left( \boldsymbol{\theta} | \mathcal{M} \right)  d\boldsymbol{\theta}$ is equal to $1$, hence $\log Z(0) =0$. 
Since the PTMCMC algorithm samples the tempered posterior distribution of $\boldsymbol{\theta}$ for all temperatures, we can estimate Eq.~(\ref{eq:log_like_exp}) by calculating the sample expectation 
\begin{eqnarray}
\mathrm{E}_{\beta}\left[ \log p\left( \boldsymbol{y} | \boldsymbol{\theta} , \mathcal{M} \right) \right] \approx \frac{1}{K}\sum_{k=1}^{K} \log p\left( \boldsymbol{y} | \boldsymbol{\theta}_{k} , \mathcal{M} \right),
\end{eqnarray}
where $K$ is the number of samples.
The integral in Eq.~(\ref{eq:evidence_integral}) in then evaluated numerically using the trapezoidal method.

We also have to estimate the error made on the evidence $Z(1)$ when using this approximation. The bias due to the integral approximation is usually larger than the variance. To evaluate it, we use the approach adopted in the \textsc{ptemcee} code~\cite{Vousden2015}, where the numerical integration is first performed on a subset of the temperature ladder, and then on the entire ladder. The difference between the two values that we obtain gives an estimate of the error due to the discretization of the continuous integral, i.e. the bias on the evidence.

\begin{acknowledgments}
We would like to gratefully thank the members of Goddard Space Flight Center Gravitational Astrophysics Group for very useful discussions, and thank the members of the LISA Consortium Simulation and the LISA Data Challenge working groups for interesting inputs and exchanges. We also thank Tyson Littenberg for interesting remarks. This work is supported by an appointment to the NASA Postdoctoral Program at the Goddard Space Flight Center, administered by Universities Space Research Association under contract with NASA.
\end{acknowledgments}

\bibliography{./myreferences}

\end{document}